\documentclass[journal]{IEEEtran}
\usepackage{graphicx} %
\usepackage{amsmath}
\usepackage{epstopdf}
\usepackage{xcolor} %

\usepackage{cite}

\usepackage{subcaption}
\usepackage[font=footnotesize,labelfont=bf]{caption}

\usepackage{url} %
\usepackage{amssymb}

\title{\huge Efficient River Water Level Sensing Using Cellular CSI and Joint Space-Time Processing}
\author{
		Khawaja Fahad Masood, \IEEEmembership{Member,~IEEE,}
        Kai Wu, \IEEEmembership{Member,~IEEE,}
		Zhongqin Wang,~\IEEEmembership{Member,~IEEE,}\\
        J. Andrew Zhang,~\IEEEmembership{Senior Member,~IEEE,}
		Shu-Lin Chen,~\IEEEmembership{Senior Member,~IEEE,} 		
		and~Y. Jay Guo,~\IEEEmembership{Fellow,~IEEE
		}%
        
\vspace{-5mm}

\thanks{Manuscript received xxxx. }
\thanks{All authors are with the Global Big Data Technologies Centre and the School of Electrical and Data Engineering, University of Technology Sydney, Australia (e-mail: \{khawajafahad.masood; kai.wu; zhongqin.wang; andrew.zhang; shulin.chen; jay.guo\}@uts.edu.au).}
}  
\begin{document}

\maketitle
\begin{abstract}
Accurate and timely water level monitoring is critical for flood prevention, environmental management, and emerging smart infrastructure systems. Traditional water sensing methods often rely on dedicated sensors, which can be costly to deploy and difficult to maintain and are vulnerable to damage during floods.%
In this work, we propose a novel cellular signal-based sensing scheme that passively estimates water level changes using downlink mobile signals from existing communication infrastructure. By capturing subtle variations in channel state information (CSI), the proposed method estimates %
the length changes of the water-reflected signal path, which correspond to water level variations. %
A space-time processing framework is developed to jointly estimate the angle of arrival and Doppler shift, enabling isolation and enhancement of the water-reflected path via beamforming, while effectively suppressing environmental noise. The phase evolution of the beamformed signal is then extracted to infer water level changes. To address clock asynchronism between the transmitter and receiver inherent in bi-static systems, we introduce a beamforming-based compensation technique for removing time-varying random phase offsets in CSI. Field experiments conducted across a river demonstrate that the proposed method enables accurate and reliable water level estimation, achieving a mean accuracy ranging from 1.5 cm to
3.05 cm across different receiver configurations and deployments.%
\end{abstract}


\section{Introduction}
The Internet of Things (IoT) has enabled unprecedented opportunities for environmental monitoring by connecting large networks of sensors, devices, and systems capable of real-time data collection and intelligent decision-making \cite{fang2014integrated, 10664494, 10210671}. However, traditional IoT deployments, especially for flood and hydrological monitoring, face critical challenges in scalability, cost, resilience to harsh environments, and maintenance. In flood-prone regions,%
conventional water-level sensors%
systems fail due to power loss, physical damage, or communication disruptions \cite{natividad2018flood}. As such, there is a growing need for resilient %
and widely accessible alternatives for environmental sensing.

Recent advances in integrated sensing and communication (ISAC), also known as joint communication and sensing (JCAS), offer a compelling solution to this problem. By leveraging existing wireless communication infrastructure, such as %
mobile networks and WIFI, ISAC enables opportunistic sensing without deploying dedicated sensors, thereby reducing cost, improving coverage, and enhancing robustness \cite{10012421, 10540249, wu2022joint}. Among various ISAC architectures, bi-static sensing (BSS) has gained significant interest due to its ability to reuse existing infrastructure. In BSS, a physically separated transmitter (Tx) and receiver (Rx) pair leverage wireless signals impacted by the environment for sensing, without interrupting ongoing communication. Technologies such as Wi-Fi, LoRa, and mobile networks have been exploited in BSS applications for intrusion detection, human tracking, gesture recognition, and environmental monitoring \cite{kosba2012rasid,xiao2013pilot,qian2018widar2,zheng2019zero,li2022csi,han2023centitrack,chen2020robust}.

A core challenge in BSS systems is clock asynchronism between Tx and Rx \cite{wu2024sensing}, which leads to time-varying random phase offsets (RPOs) arising from carrier frequency offset (CFO) and timing offset (TO). These impairments degrade the performance of fine-grained channel state information (CSI)-based sensing algorithms, especially in the Doppler and delay domains. While methods such as cross-antenna cross-correlation (CACC) \cite{qian2018widar2} and cross-antenna signal ratio (CASR) \cite{li2022csi} exist for RPO mitigation, they either suffer from %
Doppler ambiguity or are sensitive to noise \cite{wu2024sensing}. Alternative solutions reconstruct static-line-of-sight (LOS) references to estimate and correct RPOs \cite{zhao2023multiple,wu2023low}.

Integrating sensing into mobile networks leads to preceptive mobile networks (PMN) \cite{9585321}. %
PMN offers greater potential for outdoor sensing due to its broader coverage and established infrastructure. Previous PMN-based outdoor sensing efforts include %
traffic control \cite{feng2021lte}, and soil moisture sensing \cite{feng2022lte}. However, PMN’s full potential in dynamic outdoor environments, particularly for flood sensing and hydrological monitoring, remains underexplored. 
Amid rising climate risks, accurate and timely water level monitoring is essential to mitigate flood impacts \cite{yu2017comparison}. Conventional solutions using %
dedicated sensor networks, often rely on vulnerable infrastructure and face limitations under extreme conditions,
including exposure-induced failures, connectivity breakdowns, and costly maintenance demands
\cite{li2016new,natividad2018flood}. PMN-based sensing presents a promising alternative by leveraging ubiquitous infrastructure and robust signal propagation characteristics.

\begin{figure*}[!t]
\centering
  \includegraphics[width=0.8\textwidth]{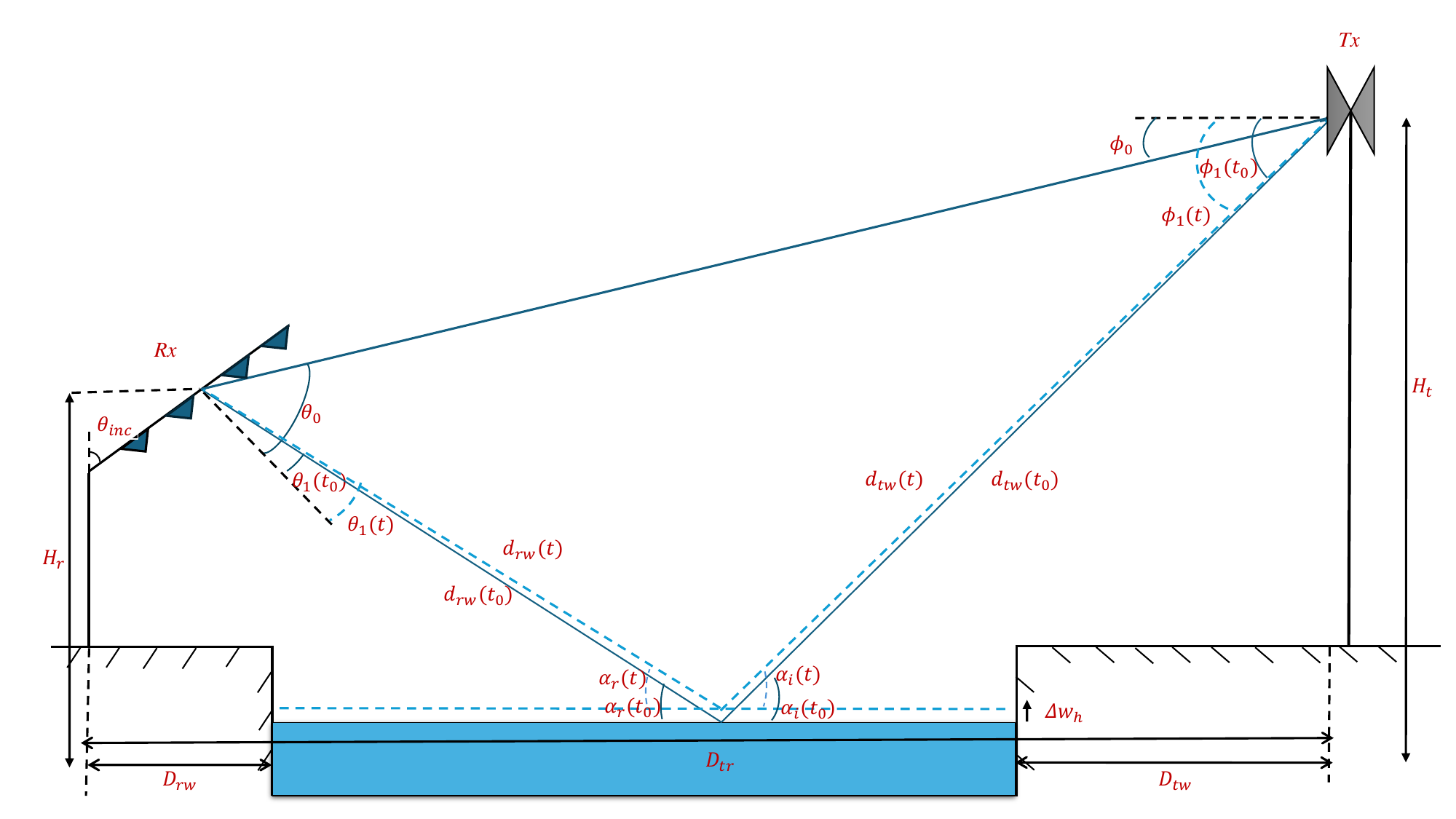} 
  \caption{System model for downlink LTE-based water level sensing. The setup includes a base station (Tx, on the right), a vertically arranged $M$-element uniform linear array (ULA) receiver (Rx, on the left), and a water reservoir (river) situated between them. The system uses the Rx array to receive cellular downlink signals, from which CSI is extracted and subsequently %
  processed by the proposed scheme for high-accuracy water level sensing.}
  \label{fig:SysM}
\end{figure*}

In this paper, we investigate the scenario illustrated in Fig.~\ref{fig:SysM}, where a downlink base station transmits signals across a water body to a limited-size vertical antenna array. Our goal is to estimate the water level by extracting and enhancing the water-reflected path from the received downlink CSI, despite the presence of multipath interference, limited aperture, and array imperfections.
Our approach includes two major components: CSI preprocessing and water level sensing. 

First, we propose a novel CSI preprocessing scheme that includes a robust RPO estimation method via beamforming. The CSI is then cleaned and transformed via a custom dimension reduction method that suppresses non-water-related paths in the delay and Doppler domains. This results in enhanced observability of the water-reflected path.

Next, %
preprocessed CSI is exploited for the proposed water level sensing scheme. We apply a joint spatial-temporal processing algorithm to isolate the water-reflected path. The phase evolution of this path reveals the change in path length due to water level variation. Using angle-of-arrival (AoA) estimation and geometric relationships, we convert the path length change into a water level estimate. The proposed scheme also accommodates array imperfection and environmental variation.

Our main contributions are summarized as follows. 
\begin{itemize}
  \item We propose a CSI preprocessing framework that includes robust RPO estimation, phase compensation, and Doppler-delay domain dimension reduction. This preprossing framework removes the phase ambiguity in Doppler and delay estimation effectively and reduces the computational effort for the proposed water level sensing scheme.%
  \item A novel water level sensing scheme is developed using joint space-time processing, beamforming-based phase tracking, and geometric inference. The proposed scheme enhances the water-reflected signal to estimate the phase change caused by water level variation.
  \item We derive the water level change from the Doppler frequency and phase evolution of the extracted path, accounting for array configuration and AoA information.
  \item The proposed scheme is validated using real-world LTE signals collected across the Paramatta River in Sydney, NSW, Australia, demonstrating reliable and accurate water level tracking, with a mean estimation error ranging from 1.5 cm to 3.05 cm over approximately 1 m of water level variation.%
\end{itemize}
The rest of the paper is structured as follows. Section II presents the system and channel model. Section III details the CSI cleaning and dimension reduction scheme. Section IV describes the proposed water level sensing algorithm. Experimental results are presented in Section V. Finally, conclusions are drawn in Section VI.

\section{System Model}

Our proposed scheme can be applied to mobile networks of different generations. Since our experiments were mainly based on LTE 4G signals due to its wide availability, we describe our scheme with respect to LTE. {Fig. \ref{fig:SysM} illustrates the system model. It consists of a base station (BS) as the transmitter (Tx), a receiver (Rx) array and a river in between them. LTE signals are transmitted by the BS, and the cell-specific reference signal (CRS) received at the Rx are used to estimate the downlink CSI for water sensing. The Rx consists of an $M$-element uniform linear array (ULA) arranged vertically to estimate the elevation angles. %
For clarity of analysis, we model the channel using two dominant components, including LOS and water-reflected paths. Nonetheless, real-world mobile signal
propagation involves more complex multipath, environmental noise, and temporal dynamics. Our method maintains reliable performance under practical conditions. Two different water levels create two different water-reflected paths. The initial water level at time $t_0$ is shown with a solid line and the water level at time $t$ is shown with a dotted line. Similarly, the paths for both water levels are also shown with solid and dotted lines, respectively. The detailed parameters are listed below.} 

In Fig. \ref{fig:SysM}, $D_{tr}$, $D_{rw}$, and $D_{tw}$ denote the distance between Tx and Rx, Rx and water, and Tx and water, respectively. Let $t$ be the time of parameter estimation and $t=t_0$ be the reference time. We are interested in estimating how  signal and water parameters change from $t_0$ to $t$. %
The angles $\theta_{inc}$, $\theta_0$ ($\phi_0$), $\theta_1\left(t\right)$ ($\phi_1\left(t\right)$) and $\phi_1\left(t\right)$ ($\phi_1\left(t\right)$), denote {the} Rx {array} inclination angle, the LOS AoA (angle-of-departure (AoD)),  water-reflected path AoA (AoD) and  at time $t$, respectively. Furthermore, $H_r$, $H_t$, $H_w\left(t\right)$ denote the heights of the center of the Rx array, the Tx, and the water surface at time $t$, respectively.
{The path length between the transmitter and water surface is given by $d_{tw}\left(t\right)$ and that between the receiver and the water surface is indicated by $d_{rw}\left(t\right)$ and  at time $t$, respectively.} The signal propagation change, as caused by an increase in the water level, is depicted in Fig. \ref{fig:SysM}.
\begin{figure}[t]
\centering
  \includegraphics[width=\columnwidth]{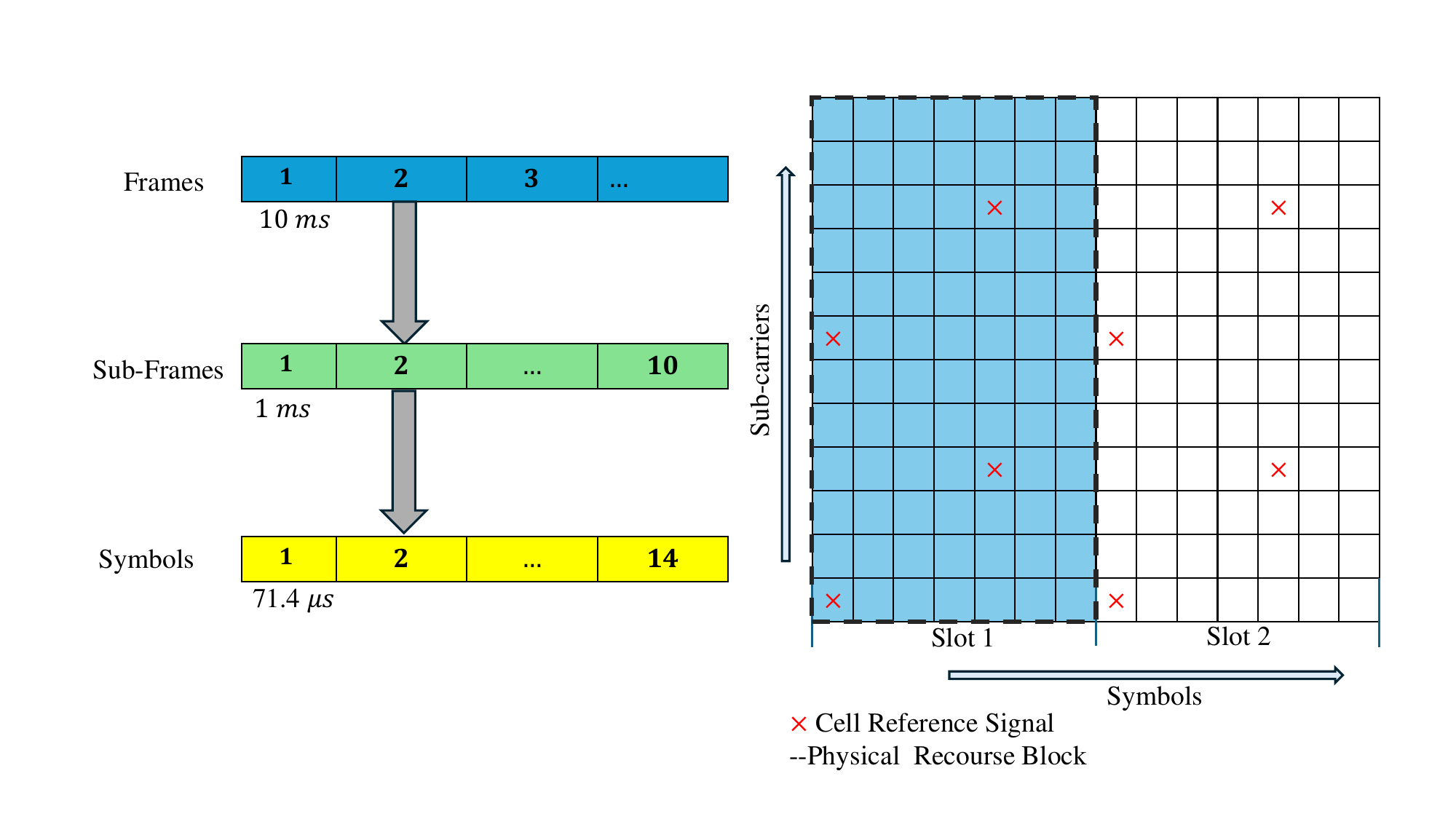} 
  \caption{LTE Frame Structure and Physical Resource blocks with CRS locations.}
  \label{fig:LTE}
\end{figure}

Based on the above system setup, we aim to use the downlink cellular signal CSI to achieve high-accuracy water level
sensing. %
Next, we illustrate the LTE signal source first and then the propagation channel, laying the theoretical foundation for the method development in the following sections.%
\subsection{LTE Frame Structure}
{In LTE, the transmission is carried out in a frame-by-frame manner, {as shown in Fig. \ref{fig:LTE}.} Each FDD-LTE\footnote{{At the authors' location, frequency division duplex (FDD) is the de-facto mechanism for LTE.}} frame has a duration of 10 $ms$. Each frame is divided into $10$ sub-frames each having $1$ $ms$ duration. Furthermore, each sub-frame has $14$ symbols with symbol duration being $71.429$ $\mu s$ approximately under the normal cyclic prefix (CP) configuration. LTE adopts orthogonal frequency division multiple access (OFDMA) in the physical layer with a sub-carrier spacing of $15$ kHz. 
The basic transmission block is called a physical resource block (PRB), as illustrated in Fig. \ref{fig:LTE}. Each PRB consists of half sub-frame (a slot with $7$ symbols) and $12$ sub-carriers. Each PRB contains $84$ symbol-sub-carrier pairs called resource elements (REs). For channel estimation, pilot symbols or %
CRSs are transmitted at known RE locations within each PRB. The location of CRSs within a PRB is determined by the physical cell identity (CID) of a cell. In LTE systems, CRS are transmitted continuously by BS even without data communications \cite{innovations2010lte}.}

{At the receiver, {CID can be obtained based on the synchronization signals}. The CRSs' indices are then derived based on CID \cite{sesia2011lte}. The known CRSs can be then used for downlink CSI estimation. In each LTE sub-frame, two
OFDM symbols are dedicated to CRS transmission, resulting
in a temporal sampling rate of 2000 Hz (i.e., one CRS-bearing
symbol every 0.5 ms). In the frequency domain, CRS symbols
are inserted every 6 subcarriers, corresponding to a spacing
of 90 kHz. This high-resolution sampling in both time and
frequency enables the reconstruction of fine-grained channel
responses. %

\subsection{Channel Model}
The downlink CSI %
for the $k$-th sub-carrier, $l$-th symbol and $m$-the receive antenna can be expressed as
\begin{equation}
\label{hklm}
{h}\left(k,l,m\right)= e^{j \mathbf{\phi}_{k,l}}\sum^{P-1}_{p=0} {h}_p\left(k,l,m\right)+n\left(k,l,m\right),
\end{equation}
where $p$ is the signal propagation path index, and the total number of paths is represented by $P$. {In (\ref{hklm}), $\phi_{k,l}=\phi^f_{k,l}+\phi^\tau_{k,l}+\phi^i_{k,l}$ denotes the RPO which is the combined effect of CFO, $\phi^f_{k,l}$, TO, $\phi^{\tau}_{k,l}$, and a random initial phase, $\phi^i_{k,l}$.}Moreover, $n\left(k,l,m\right)$ denotes the complex independent and identically distributed (i.i.d.) Gaussian noise. The RPO is the same for all the receive antennas and vary over sub-carriers and symbols. The channel for the $p$-th path can be further represented as
\begin{equation}
  {h}_p\left(k,l,m\right)=\alpha_p e^{-j2\pi \left(f_k \tau_p\right)}e^{j2\pi \left(t_l f_{d_p}\right)} e^{-j2\pi \left(\kappa_m\sin\left(\theta_p\right)/\lambda\right)},
\end{equation}
where $\alpha_p$, $\tau_p$, $f_{d_p}$, and $\theta_p$ denote the complex path gain, delay, Doppler frequency shift (DFS) and AoA for the $p$-th path, respectively. Moreover, $f_k=f_c+\Delta_f \left(k-1\right)$, $t_l=\Delta_t \left(l-1\right)$, and $\kappa_m=\kappa\left(m-1\right)$ represent the frequency of the $k$-th sub-carrier, the time of the $l$-th symbol, and the distance spacing of the $m$-th antenna from the reference antenna, respectively. Furthermore, $\Delta_f$, $\Delta_t$, and $\kappa$ represent the sub-carrier spacing, the symbol spacing, and the antenna spacing, respectively. 
For convenience of further analysis, we introduce the downlink channel matrix for the $k$-th sub-carrier, as given {by
\begin{align}
    \mathbf{H}_k=& \left(\sum^{P-1}_{p=0} \alpha_p e^{-j2\pi \left(f_k  \tau_p\right)}\mathbf{a}\left(\theta_p\right)\mathbf{d}_p^{\text{T}}\right)\text{diag}\left(\mathbf{\Phi}_{k}\right) +\mathbf{N}_{k},%
      \label{ksubH}
\end{align}
where $\text{diag}\{\cdot\}$ generates a diagonal matrix based on the enclosed vector, $\mathbf{\Phi}_{k}=\left[e^{j\phi_{k,0}},e^{j\phi_{k,2}},\dots, e^{j\phi_{k,L-1}}\right]$ represents the RPO vector for the $k$-th sub-carrier, and $\mathbf{N}_k$ denotes the noise term for the the $k$-th sub-carrier's CSI. In (\ref{ksubH}), $\mathbf{a}\left(\theta_p\right)$ is the steering vector for the $M$-element Rx ULA. It can be given by 
\begin{align}
  \mathbf{a}\left(\theta_p\right)= \left[1, e^{-j2\pi \left(\kappa\sin\left(\theta_p\right)/\lambda\right)}, \dots, e^{-j2\pi \left(\kappa\left(M-1\right)\sin\left(\theta_p\right)/\lambda\right)}\right]^\text{T}.
  \label{Steering}
\end{align}
 Moreover, $\mathbf{d}_p$ denotes the Doppler response vector for $L$ symbols relative to the initial symbol, as given by
 \begin{align}
  \mathbf{d}_p= \left[1, e^{j2\pi \left(\Delta_t f_{d_p}\right)}, \dots, e^{j2\pi \left(\Delta_t\left(L-1\right) f_{d_p}\right)}\right]^{\text{T}}.
  \label{Steering}
\end{align}
}

From (\ref{ksubH}), we see that the columns of $\mathbf{H}_k$ are each multiplied by a different unknown RPO, i.e., a diagonal element of $\mathbf{\Phi}_{k}$. Moreover, $\mathbf{\Phi}_{k}$ changes over sub-carriers, as indicated by its subscript $k$. 
{These prevent us from using the raw CSI signals for direct water level sensing. Therefore, the proposed water level sensing scheme starts with a RPO compensation method. It then combines CSI symbols over time to estimate the Doppler frequency caused by water level changing. The scheme also includes a method to convert the Doppler estimation to actual water level. 
To reduce computational and storage needs, a CSI dimension reduction approach is also proposed for the water sensing scheme. The details of different components in the scheme are presented in the next section.}

\section{CSI cleaning and Dimension Reduction}
{In this section, we develop methods for RPO estimation, CSI cleansing and dimension reduction. Fig. \ref{fig:CSIcleaning} shows an overview of the whole preprocessing scheme. We propose an optimal beamforming-based approach to estimate the RPO. %
Then the CSI is cleaned by compensating the RPO. Next, the CSI dimension reduction scheme is proposed; the proposed CSI dimension reduction not only reduces the CSI dimension but also enhances the water-reflected signal by suppressing unwanted paths in delay and Doppler domains. 
}

 \begin{figure*}[ht]
\centering
  \includegraphics[width=0.8\textwidth]{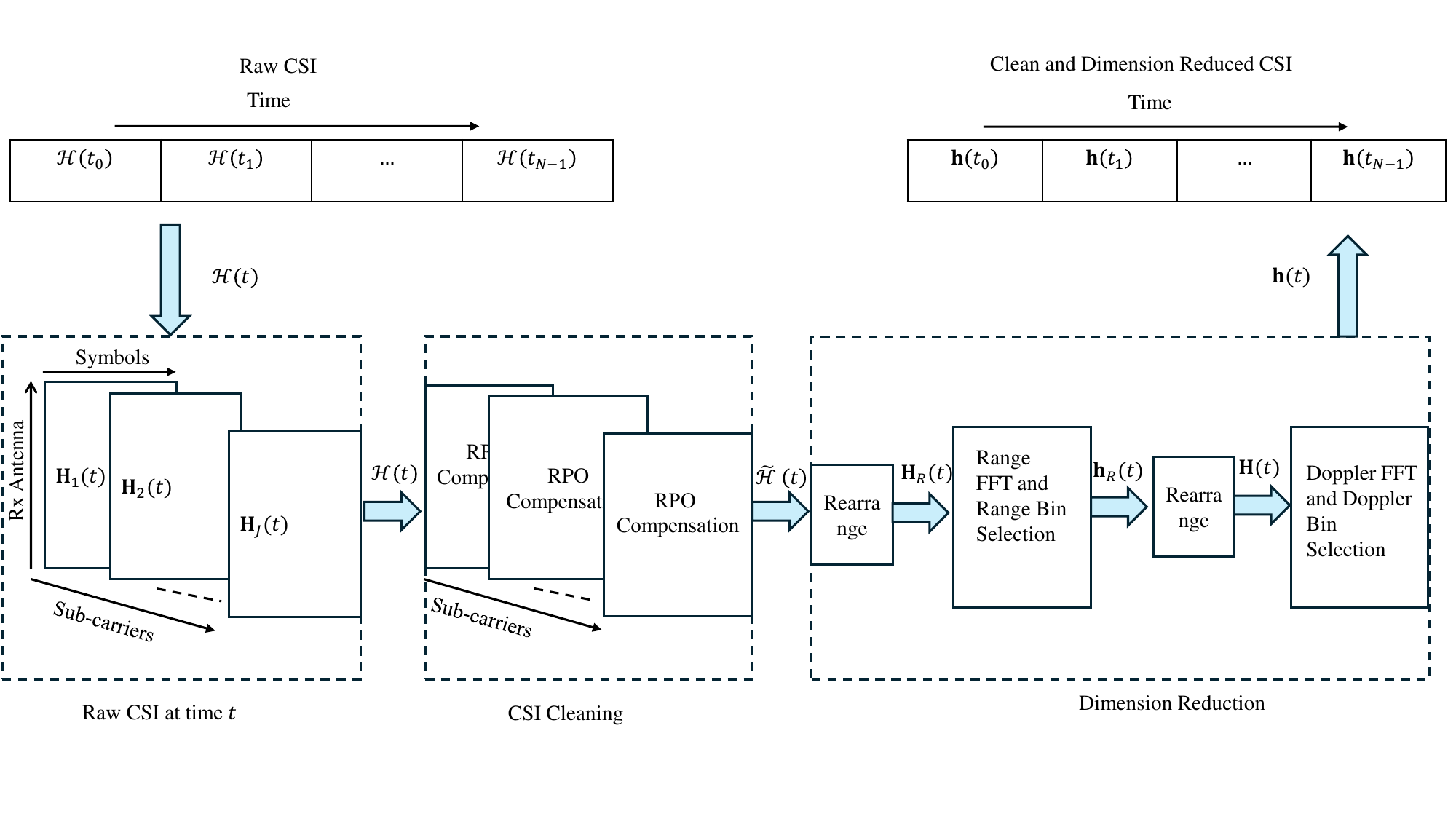} 
  \caption{The proposed CSI preprocessing scheme, including cleaning and dimension reduction, for high-accuracy water level sensing.}
  \label{fig:CSIcleaning}
\end{figure*}
\subsection{Proposed Random Phase Offset Compensation}
\label{sec:RPO}
{Clock asynchronism is a known challenge in bi-static ISAC systems \cite{wu2024sensing}}. Two major approaches used for RPO compensation include the CACC and CASR. Both of these strategies exploit the fact that the RPO are the same over different antennas and use the conjugate multiplication/division operation of a reference antenna CSI with remaining antennas. However, CACC is prone to sign ambiguity, which may lead to incorrect Doppler/delay estimates. CASR division can cause noise amplification if the reference antenna signal is of small magnitude. Moreover, taking the ratio in CASR can lead to complicated signal models, invalidating conventional signal processing methods.
Since the scenario of interest has the presence of LOS\footnote{If LoS is not present, a strong static path can be identified from the deployment scenario for the proposed scheme.}, we consider using LOS to construct a reference signal to mitigate RPO on other paths.
For clarity of illustration, two dominant paths\footnote{{The proposed RPO compensation approach is readily extendable to scenarios with more than two paths (static or dynamic paths), as long as a strong static path is present.}} are considered for now, with $p=0$ as the LOS path and $p=1$ as the water-reflected path. The channel model given in (\ref{ksubH}) can be rewritten as
\begin{align}
    \mathbf{H_k}=& \left(\mathbf{H}_{k_{0}}+\mathbf{H}_{k_{1}}\right)\cdot\text{diag}\left(\mathbf{\Phi}_{k}\right)+\mathbf{N}_{k} \nonumber\\
        =& \left(\mathbf{h}_{k_{0}}\cdot\mathbf{1}+\mathbf{H}_{k_{1}}\right)\cdot\text{diag}\left(\mathbf{\Phi}_{k}\right)+\mathbf{N}_{k}.
\end{align}
{Since the LOS path is static, $\mathbf{d}_0=\mathbf{1} \in \mathbb{C}^{1\times L}$. Based on (\ref{ksubH}), $\mathbf{h}_{k_0}=\alpha_0 \mathbf e^{-j2\pi \left(f_k  \tau_0\right)}{\mathbf{a}}\left(\theta_0\right) \in \mathbb{C}^{M \times 1}$. Then the temporal variation in LOS path mainly comes from the RPO, mathematically}
\begin{align}
\mathbf{H_k}=& \mathbf{h}_{k_{0}}\cdot\mathbf{1}\cdot\text{diag}\left(\mathbf{\Phi}_{k}\right)+\mathbf{H}_{k_{1}}\cdot\text{diag}\left(\mathbf{\Phi}_{k}\right)+\mathbf{N}_{k} \nonumber\\
=& \mathbf{h}_{k_{0}}\cdot{\mathbf{\Phi}}_{k}+\mathbf{H}_{k_{1}}\cdot\text{diag}\left(\mathbf{\Phi}_{k}\right)+\mathbf{N}_{k},
\end{align}
{where we recall that $\mathbf{\Phi}_{k}$ denotes the vector of RPO over the symbols.}
To extract the signals over the LOS path, we utilize the minimum variance distortionless response (MVDR) beamformer, an optimum beamformer. It starts with calculating the sample covariance matrix of $\mathbf{H}_k$, as given by
 \begin{align}
   \mathbf{R}_k=\frac{1}{L}\mathbf{H}_k\mathbf{H}_k^H +\rho \mathbf{I},
 \end{align}
 where $\rho$ is the diagonal loading factor to ensure the covariance matrix is well conditioned. Then the weights of the MVDR beamformer are calculated by solving the following optimization problem
 \begin{align}
\min_{\mathbf{w}} \mathbf{w}^H \mathbf{R}_k \mathbf{w} \ \ \text{subject to} \ \
\mathbf{w}^H \mathbf{a}\left(\theta_0\right) = 1,
\label{MVDRBF1}
\end{align}
 with a closed-form solution \cite{van2002optimum},
\begin{align}
  \mathbf{w} = \frac{\mathbf{R}_k^{-1} \mathbf{a}\left(\theta_0\right)}{\mathbf{a}^H\left(\theta_0\right) \mathbf{R}_k^{-1} \mathbf{a}\left(\theta_0\right)},
\end{align}
 where $\left(.\right) ^{-1}$ takes the inverse of a matrix and $\mathbf{a}\left(\theta_0\right)$ is the steering vectors towards the LOS path. Then using these weights, a reference signal is constructed, as given by
 \begin{align}
   \mathbf{r_k}=&\mathbf{w}^H\mathbf{H}_k  
     ={h}_{k_{0}}.\mathbf{\Phi}_{k} +\mathbf{e} 
     \approx {h}_{k_{0}}.\mathbf{\Phi}_{k},
 \end{align}
 where $\mathbf{e}$ represents the collective effect of noise and the errors due to other paths. Then, the phase variations over the symbols in the reconstructed reference signal can be regarded as the RPO. Finally, the RPO are compensated using the estimated phase of the reconstructed reference signal as shown below, 
\begin{align}
          \mathbf{\tilde{H}_k}=&\mathbf{{H}_k} .\text{diag}\left({e^{-j \angle\left(\mathbf{r_k}\right)}}\right) 
          =\left(\mathbf{h}_{k_{0}}.\mathbf{1}+\mathbf{H}_{k_{1}}\right) .{e^{j\phi_r}}+\mathbf{N}_{k},
\end{align}
where $\mathbf{\tilde{H}_k}$ denotes the clean CSI for the $k$-th sub-carrier after RPO compensation. {Also, $\phi_r=-\angle\left(\mathbf{w}^H \mathbf{h}_{k,0}\right)$ is the phase offset residue. This residue is time-invariant and can be shown as rotation in the compensated CSI. Due to the time-invariance of the residue, the Doppler estimation is not corrupted by phase offset residue.} This same process of RPO compensation {can be applied to all sub-carriers.}%

The performance of the RPO compensation scheme is shown in Fig. \ref{fig:CFOComp}. {The results are obtained based on data collected in field experiments for Setup 1 in Section \ref{sec:FieldExp}. With an experimental scenario similar to the illustration in Fig. \ref{fig:DataCollectionSetup}, more details of the experiments will be provided in Section \ref{sec:FieldExp}}. The AoA of the LOS path is $47$ degrees, and since both Tx and Rx are stationary, the DFS should be zero. However, it can be seen in Fig. \ref{fig:CFOComp}(a) that raw CSI introduces an ambiguity in the Doppler domain. However, in Fig. \ref{fig:CFOComp}(b), after RPO compensation, the ambiguity is removed and the LOS path can be seen with zero DFS.
\begin{figure}[!t] %
  \centering %
  \begin{subfigure}{0.4\textwidth} %
    \centering
    \includegraphics[width=\linewidth]{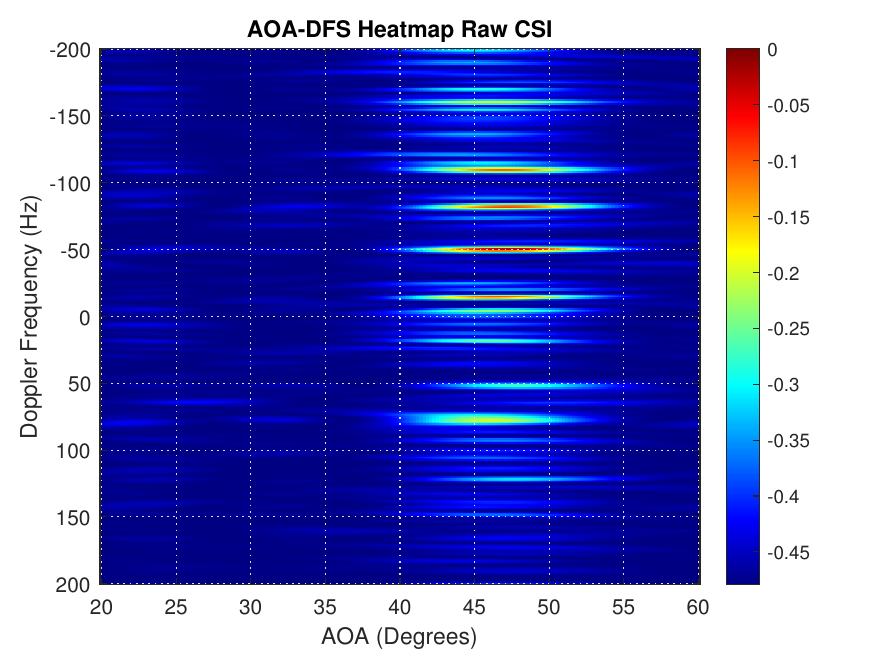} %
    \caption{Before RPO Compensation.} %
    \label{fig:subfig1} %
  \end{subfigure}
  \begin{subfigure}{0.4\textwidth} %
    \centering
    \includegraphics[width=\linewidth]{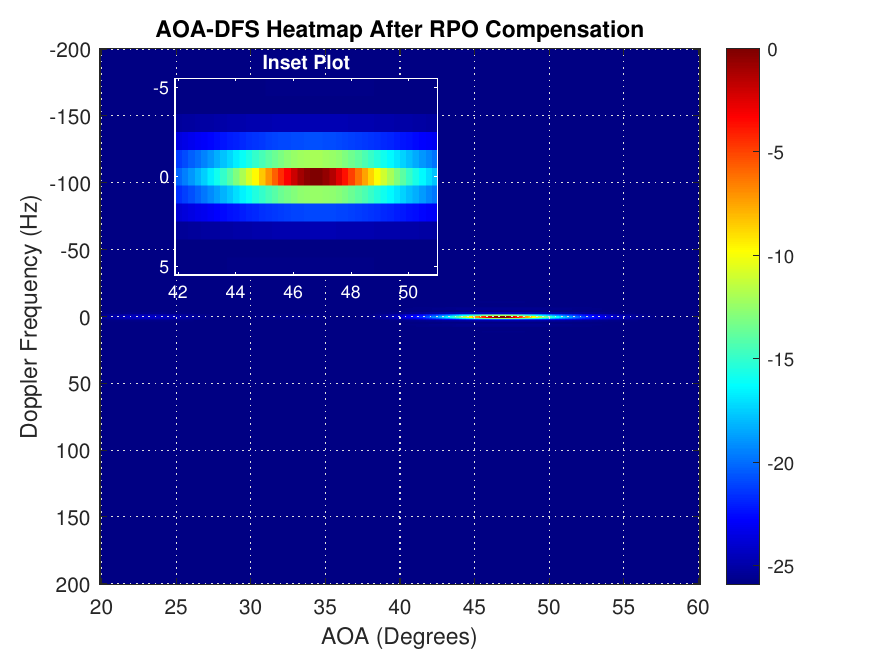} %
    \caption{After RPO Compensation.} %
    \label{fig:subfig2} %
  \end{subfigure}
  \caption{Doppler Frequency for LOS path before and after RPO compensation.} %
  \label{fig:CFOComp} %
\end{figure}

\subsection{Proposed CSI Dimension Reduction}
\label{sec:DimRed}

Each CSI capture is a three-dimensional (3D) matrix, denoted by ${\mathcal{H}}\left(t\right)$, with dimensions $K \times L \times M$. For each CSI capture ${\mathcal{H}}\left(t\right)$, the delay and Doppler domains have limited resources due to the small bandwidth and short CSI capture time, respectively. Therefore, we propose to estimate the water-reflected path using the spatial domain. Then it is convenient to reduce the dimensions in the delay and Doppler domains with signal quality enhancement in the meantime.

The CSI dimension reduction flow diagram is depicted in the bottom right of Fig. \ref{fig:CSIcleaning}. {In the first step, 
we 
convert the frequency-domain signals, i.e., over sub-carriers, to the range domain, as detailed below.} First, the CSI matrix $\mathcal{H}\left(t\right)$ is rearranged to form a two-dimensional (2D) matrix $\mathbf{H}_R\left(t\right)$ with dimensions $K\times (L\times M)$, concatenating the CSI of all antennas. Then, a fast Fourier transform (FFT) is applied to each column of $\mathbf{H}_R\left(t\right)$ to reduce the number {of} sub-carriers $K$ to the number of range bins, $R_b$. For the considered scenario, i.e. LOS with both Tx and Rx facing the water, both the LOS and water-reflected paths are captured in the strongest range bin, since the path difference between the LOS path and water-reflected is less than the range resolution. Therefore, the strongest range bin is selected, as it will contain both paths. Those paths that do not fall into the selected bin are suppressed. This reduces $\mathbf{H}_R\left(t\right)$ to a vector $\mathbf{h}_R\left(t\right)$ with dimensions $LM\times 1$.  

In the second step, as shown in Fig. \ref{fig:CSIcleaning}, the vector $\mathbf{h}_R\left(t\right)$ is rearranged as a matrix $\mathbf{H}\left(t\right)$ with dimensions $L\times M$. The water-reflected path is assumed to be static as the CSI capture duration is very small, e.g., $0.2$s. Thus, the strongest Doppler bin is selected, which contains the LOS path, the water-reflected path, and other static paths. This also helps suppresses the interference introduced by fast-moving objects, in contrast to the slow varying water level changes. 
This reduces $\mathbf{H}\left(t\right)$ to $\mathbf{h}\left(t\right)$ with dimensions $M\times 1$, i.e. a single value for each antenna. This dimension reduction is applied on all the CSI captures. Therefore, the resulting matrix is denoted by $\mathbf{H}$ with dimensions $M\times N$, where $N$ is the total number of CSI captures. {Using the concatenated channel matrix $\mathbf{H}$ obtained in this section, the proposed water level sensing scheme is presented next.}

 \section{Water Level Sensing Scheme}

{The proposed scheme exploits the variations in the CSI over time to estimate the water level change. The flow diagram of the scheme is depicted in Fig. \ref{fig:FlowDiagram}, with inputs from the cleaned and dimension-reduced CSI obtained in Section III. The water level change introduces a change in the water-reflected path length. This change in water-reflected path length causes a phase variation in water-reflected path's signal. To extract the phase variation, the water-reflected signal is first enhanced by suppressing interfering paths via identifying its AoA based on a joint space-time processing scheme. With the estimated AoA, %
the signal from water-reflected path is then enhanced by applying high-resolution spatial and temporal filtering. %
Finally, the water level change is derived via estimating the DFS from the enhanced water-reflected signal.
\begin{figure*}[ht]
\centering
  \includegraphics[width=0.8\textwidth]{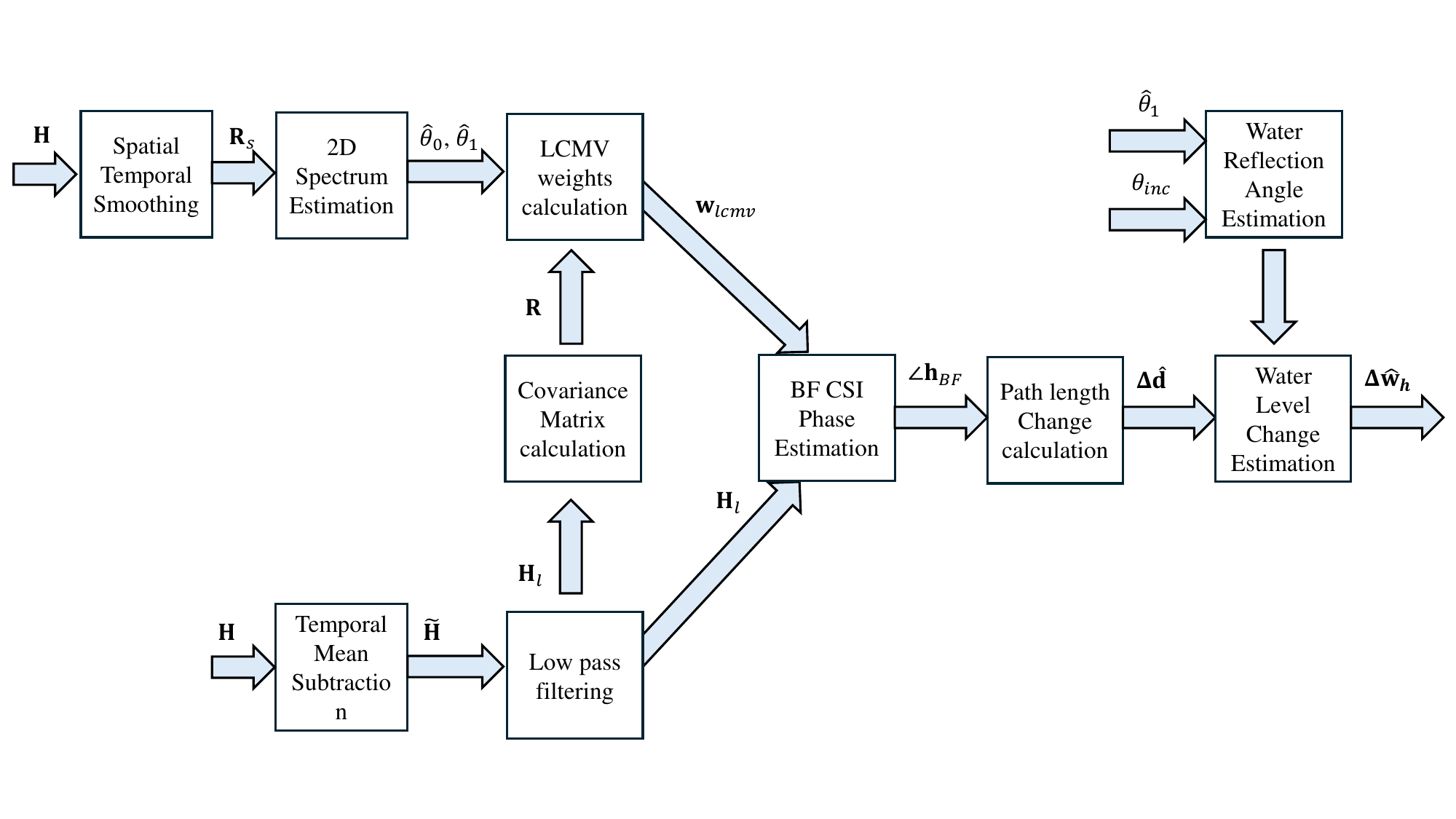} 
  \vspace{-5mm}
  \caption{The Proposed Water level Sensing Scheme.}
  \label{fig:FlowDiagram}
  \vspace{-5mm}
\end{figure*}}

\subsection{Water-Reflected Path AoA Estimation}
\label{sec:AoAEstimate}
{In Fig. \ref{fig:FlowDiagram}, spatial-temporal smoothing and 2D spectrum estimation can be seen as the initial two steps.}
After the dimension reduction the CSI over the total time $T$ can be written as
\begin{align}
      \mathbf{H}=\left[\mathbf{h}\left(t_0\right), \mathbf{h}\left(t_1\right), \dots, \mathbf{h}\left(t_{N-1}\right)\right] \in \mathbb{C}^{M \times N}
\label{concat_H},
\end{align} 
where $N$ is the total number of CSI captures over the total time $T$. 
The change in water level causes a change in the length of the water-reflected path. {Thus we can estimate the path length change to deduce the change in the water level.} This change can be {approximated by a linear model in a short time interval}
\begin{align}
  d\left(t\right)= d\left(t_0\right)+n\Delta t v,
\end{align}
where $n$ is the number of CSI captures, $d\left(t_0\right)$ {is the path length at the initial time $t_0$ and $d\left(t\right)$ is the path length at time $t$. Furthermore, $v$ is the rate of change in path length and $\Delta t$ denotes the interval between consecutive CSI captures}. %
This path length change introduces a slow-time Doppler effect. This slow-time Doppler effect is caused by the water body moving toward or away from the Rx as the water level increases or decreases, respectively. Therefore, to identify the AoA of the water-reflected path, 2D spectrum estimation can be used to provide the AoA and slow-time DFS to enhance the resolution of both dimensions.%

After the dimension reduction in the previous section, the channel vector at time $t_0$ can be written as
\begin{align}
\label{h_t0}
  \mathbf{h}\left({t_0}\right)= & \mathbf{h}_0+\mathbf{h}_1\left(t_0\right)+\mathbf{n}\left(t_0\right) \nonumber
  \\
  = &\mathbf{a}\left(\theta_0\right)\alpha_0 e^{-j2\pi \left(f_c \tau_0\right)} +\\&\mathbf{a}\left(\theta_1\left(t_0\right)\right)\alpha_1\left(t_0\right) e^{-j2\pi \left(f_c \tau_1\left(t_0\right)\right)} \nonumber +\mathbf{n}\left(t_0\right),
\end{align}
where $\mathbf{h}_0=\mathbf{a}\left(\theta_0\right)\alpha_0 e^{-j2\pi \left(f_c \tau_0\right)}$ is the LOS path and does not change over time, $\mathbf{h}_1\left(t_0\right)=\mathbf{a}\left(\theta_1\left(t_0\right)\right)\alpha_1\left(t_0\right) e^{-j2\pi \left(f_c \tau_1\left(t_0\right)\right)} $ is the water-reflected path at time $t_0$, and $\mathbf{n}\left(t_0\right)$ denotes the noise at time $t_0$. The time delay can be written in terms of path length as, $\tau={d}/{c}$, where $d$ is the path length, $c$ is the speed of light. %
Then we can write the water-reflected path at time $t_0$ as
\begin{align}
\mathbf{h}_1\left(t_0\right)=&\mathbf{a}\left(\theta_1\left(t_0\right)\right)\alpha_1\left(t_0\right) e^{-j2\pi \left(d_1\left(t_0\right)/\lambda\right)} 
    \label{h_1},
\end{align}
where $f_c /c$ is replaced with $1 /\lambda$. {The AoA and slow-time DFS can be estimated separately by applying spectrum estimation over the first and second dimensions of $\mathbf{H}, $ respectively.} However, to identify the water-reflected path, AoA and slow-time DFS pairing is required. Hence, the joint 2D spectrum estimation is utilized. The AoA steering vector is the same as in (\ref{Steering}). The slow-time DFS vector for the $p$-th path can be written as
\begin{align}
  \mathbf{a}\left(f_p\right)=&\left[1, e^{-j2\pi \left(d_p\left(t_1\right)/\lambda\right)},\dots,e^{-j2\pi \left(d_p\left(t_{N-1}\right)/\lambda\right)}\right], 
\end{align}
where $d_p\left(t_1\right)=d_p\left(t_0\right)+v\Delta t$ and $v_p/\lambda=f_p$. The slow-time Doppler response vector over time $T=N\Delta t$ is
\begin{align}
  \mathbf{a}\left(f_p\right)=&\left[1, e^{-j2\pi \left(\Delta t f_p\right)},\dots,e^{-j2\pi \left( \left(N-1\right)\Delta t f_p\right)}\right],
\end{align}
where $\Delta t$ denotes the sampling interval between two successive CSI captures. Then the joint 2D array can be written as
\begin{align}
  \mathbf{a}(\theta_p,f_p)=\mathbf{a}\left(f_p\right)^T \otimes \mathbf{a}\left(\theta_p\right) \in \mathbb{C}^{MN\times 1},
\end{align}
where $\otimes$ denotes the Kronecker product. Accordingly, we can vectorize the channel matrix given in (\ref{concat_H}), achieving 
\begin{align}
&\mathbf{h}=\text{vec}\left(\mathbf{H}\right) \nonumber \\
&=\left[\mathbf{h}_{1,1},\mathbf{h}_{2,1},\dots, \mathbf{h}_{M,1}, \mathbf{h}_{2,2},\dots, \mathbf{h}_{M,2}, \dots, \mathbf{h}_{M,N}\right]^T,
\label{vectH}
\end{align}
where $\text{vec}\left(.\right)$ vectorizes a matrix column-wise. Since (\ref{vectH}) represents a single snapshot and it's challenging to estimate parameters of multiple paths with a single snapshot. Spatial-temporal smoothing is applied to create multiple virtual snapshots. {For example, consider a single snapshot case with $M=4$, and $N=10$. To convert this into four snapshots having two snapshots in the spatial domain, $M_s=2$ and two snapshots in the temporal domain, $N_s=2$, the smoothed channel matrix can be rearranged into}
\begin{align}
  \mathbf{H}_s=\begin{bmatrix}
&\mathbf{h}_{1,1}, &\mathbf{h}_{2,1},&\mathbf{h}_{1,2} ,&\mathbf{h}_{2,2}\\
& \vdots & \vdots & \vdots &\vdots\\
&\mathbf{h}_{3,1}, &\mathbf{h}_{4,1}, &\mathbf{h}_{3,2}, &\mathbf{h}_{4,2} \\ 
&\mathbf{h}_{1,2}, &\mathbf{h}_{2,2}, &\mathbf{h}_{1,2}, &\mathbf{h}_{2,2}\\
& \vdots & \vdots & \vdots &\vdots\\
&\mathbf{h}_{3,2}, &\mathbf{h}_{2,2}, &\mathbf{h}_{3,2}, &\mathbf{h}_{4,2}\\ 
&\vdots &\vdots &\vdots &\vdots\\
&\mathbf{h}_{1,9}, &\mathbf{h}_{4,9}, &\mathbf{h}_{1,10}, & \mathbf{h}_{2,10}\\
& \vdots & \vdots & \vdots & \vdots\\
&\mathbf{h}_{3,9} &\mathbf{h}_{4,9},, &\mathbf{h}_{3,10}, &\mathbf{h}_{4,10} 
  \end{bmatrix}.
\end{align}
The spatial domain has two overlapping subsets, $\mathcal{M}_1 = \{1 \le m \le 3\}$ and
$\mathcal{M}_2 = \{2 \le m \le 4\}$. Similarly, the time domain has two subsets, $\mathcal{N}_1 = \{1 \le n \le 9\}$ and
$\mathcal{N}_2 = \{2 \le n \le 10\}$. In $\mathbf{H}_s$, Each column represents a snapshot. For snapshot one to four, the space-time subsets combinations used are, \{$S_1,T_1$\}, \{$S_2,T_1$\}, \{$S_1,T_2\}$ and \{$S_2,T_2\}$, respectively.
The dimensions after smoothing can be generalized as $\left(M-M_s+1\right)\left(N-N_s+1\right)\times M_s N_s$. For convenience, we introduce $\tilde{M}=M-M_s+1$ and $\tilde{N}=N-N_s+1$. Then the smoothed covariance matrix can be constructed and 2D spectrum estimation can be applied to jointly estimate the AoA and slow-time DFS as
\begin{align}
  \mathbf{R}_s= &\frac{1}{M_sN_s}\mathbf{H}_s\mathbf{H}_s^H \ \ \ \in \mathbb{C}^{\tilde{M}\tilde{N} \times\tilde{M}\tilde{N}} \nonumber \\
    = &\mathbf{A}_s\left(\theta,f\right) \mathbf{\Delta}\mathbf{A}_s\left(\theta,f\right)^H + \mathbf{\Sigma},
\end{align}
where $\mathbf{\Delta}$ denotes the source covariance matrix, and $\mathbf{\Sigma}$ represents the noise covariance matrix. Moreover, 
\begin{align}
  \mathbf{A}_s\left(\theta,f\right)=\left[\mathbf{a}_s\left(\theta_0,f_0\right),\dots, \mathbf{a}_s\left(\theta_{P-1},f_{P-1}\right) \right],
\end{align}
 is the smoothed joint response matrix, where $\mathbf{a}_s\left(\theta_p,f_p\right)=\mathbf{a}_s\left(f_p\right)^\text{T} \otimes\mathbf{a}_s\left(\theta_p\right)$ is the joint smoothed response vector for the $p$-th path. 

The above signal processing enables us to apply the 2D spectrum estimation techniques such as 2D multiple signal classification (MUSIC) algorithm \cite{schmidt1986multiple,stoica1989music} to jointly estimate the AoA-DFS. 
The eigen value decomposition (EVD) is applied on the covariance matrix $\mathbf{R}_s$, with the signal subspace represented as $\mathbf{E}_s=\left[\mathbf{e}_{1},\mathbf{e}_{2},\dots, \mathbf{e}_{P}\right]$ and the noise subspace being $\mathbf{E}_n=\left[\mathbf{e}_{P+1},\dots, \mathbf{e}_{\tilde{M}\tilde{N}}\right]$. Here $\mathbf{e}_i$ represents the $i$-th eigen vector of $\mathbf{R}_s$.
Then the 2D-MUSIC power spectrum is given as
\begin{align}
  P\left(\theta,f\right)=\frac{1}{\mathbf{a}_s\left(\theta,f\right) \mathbf{E_n}\mathbf{E_n}^H \mathbf{a}^H_s\left(\theta,f\right)},
\end{align}
with peaks formed when $\mathbf{a}_s\left({\theta,f}\right)$ matches the AoA-DFS pair of a path.  

\subsection{Extraction of Water Reflected Path}
\label{sec:WRpathExtraction}
Once the AoA of the non-line-of-sight (NLOS) path is estimated using the 2D spectrum estimation, the water-reflected signal is extracted next. 
{In Fig. \ref{fig:FlowDiagram}, this step is referred to as beamforming (BF)-based CSI phase estimation.}
{Jointly inspecting (\ref{concat_H}) and (\ref{h_t0}) show that the LOS path and other static clutter do not change over time, hence long-term temporal average can be removed to suppress the LOS path and static clutter. This can be performed based on the concatenated channel $\mathbf{H}$ given in (\ref{concat_H}), as given by}
\begin{align}
  \tilde{\mathbf{H}}=\mathbf{H}-\text{mean}\left(\mathbf{H}\right),
  \label{eq:clutter}
\end{align}
where $\tilde{\mathbf{H}}$ contains the time-varying path i.e. the water-reflected path. Due to the slow variation in the water level, the water-reflected path varies very slowly over time. The high frequency variations over time are caused by noise and unwanted changes in the surroundings. Therefore, a low-pass filter (LPF) can be used to remove the high frequency variations over time. Let $\mathbf{H}_l$ denote the filtered $\tilde{\mathbf{H}}$.

In the low signal to noise (SNR) scenarios, the water-reflected path may not yet be clean enough for water level sensing using $\mathbf{H}_l$. To improve the SNR, an optimum BF is used to further enhance the water-reflected path. Since, $\hat{\theta}_0$, and $\hat{\theta}_1$ are already estimated, linearly constrained minimum variance (LCMV) beamforming weights can be calculated. The covariance matrix for LCMV weights calculation is then written as
 \begin{align}
   \mathbf{R}=\frac{1}{N}  {\mathbf{H}_l}  {\mathbf{H}_l}^H+\tilde{\rho} \mathbf{I},
 \end{align}
where $\tilde{\rho}$ is the diagonal loading factor to ensure the covariance matrix is well conditioned. The LCMV beamformer can be designed via the following optimization problem \cite{van2002optimum},
 \begin{align}
\min_{\mathbf{w}_{lcmv}} \mathbf{w}^H_{lcmv} \mathbf{R} \mathbf{w}_{lcmv} \ \ \text{subject to} \ \
\mathbf{C}^H \mathbf{w}_{lcmv} = \mathbf{f}. 
\label{MVDRBF1}
\end{align}
where $\mathbf{f}=\left[1,0\right]^\text{T}$ is the gain vector. Then LCMV weights are calculated using the following closed form,
\begin{align}
  \mathbf{w}_{lcmv} = \mathbf{R}^{-1} \mathbf{C}\left[\mathbf{C}^H\mathbf{R}^{-1} \mathbf{C}\right]^{-1}\mathbf{f},
\label{eq:BFWeights}
\end{align}
where $\mathbf{C}=\left[\mathbf{a}\left(\hat{\theta}_1\right), \mathbf{a}\left(\hat{\theta}_0\right)\right]$ is the steering matrix. The gains indicate the desired direction $\hat{\theta}_1$ and null towards $\hat{\theta}_0$. Then BF CSI can be obtained by using 
\begin{align}
  \mathbf{h}_{BF}=\mathbf{w}_{lcmv}^H {\mathbf{H}_l}  \ \ \in\mathbb{C}^{1 \times N},
  \label{BFCSI}
\end{align}
where $\mathbf{h}_{BF}$ is termed as the BF CSI vector. The BF CSI vector, $\mathbf{h}_{BF}$ contains a dominant water-reflected path and residue from strongly suppressed remaining paths.

\subsection{Estimating Change in Water Reflected Path Length}
Once the water-reflected path is extracted, the variation in the phase of the water-reflected path can be estimated. {In Fig. \ref{fig:FlowDiagram}, this step is termed as the path length change calculation.}

{As explained earlier in (\ref{concat_H}) and (\ref{h_1}), the variation in water level causes a change in the length of water-reflected path}. This change in water-reflected path length introduces a change in the phase of the water-reflected path. However, when the distance between the Tx and Rx is large, the change in water-reflected AoA is negligible. As the distance between the Tx and Rx is usually hundreds of meters, the Tx beamwidth is usually wide to provide larger coverage. %
The Rx has a limited number of antennas, the Rx beamwidth is also wide, limiting the high-resolution AoA estimation. Then we can assume $\theta_1$ approximately remains the same as the water level slightly changes. {This assumption is validated through a simulation. Assuming a perfect reflection from the water surface, the variation in the AoA of the water-reflected path for a 1m water level change can be seen in Fig. \ref{fig:AoAvariation}. The distance between Tx and Rx is varied from $100$ m to $1000$ m. It can be seen that as the distance increases, the variation in AoA due to water level change is almost negligible. This distance is represented as $D_{tr}$ in the system model as shown in Fig. \ref{fig:SysM}. For the considered experimental setups, $D_{tr}=423$ m for Setups 1 and $D_{tr}=465$ m for Setup 2, as shown in Table \ref{tab:Setups}. For these setups, the AoA variation is around 0.2 degrees according to Fig. \ref{fig:AoAvariation}.}
\begin{figure}[bt]
\centering
\includegraphics[width=\columnwidth]{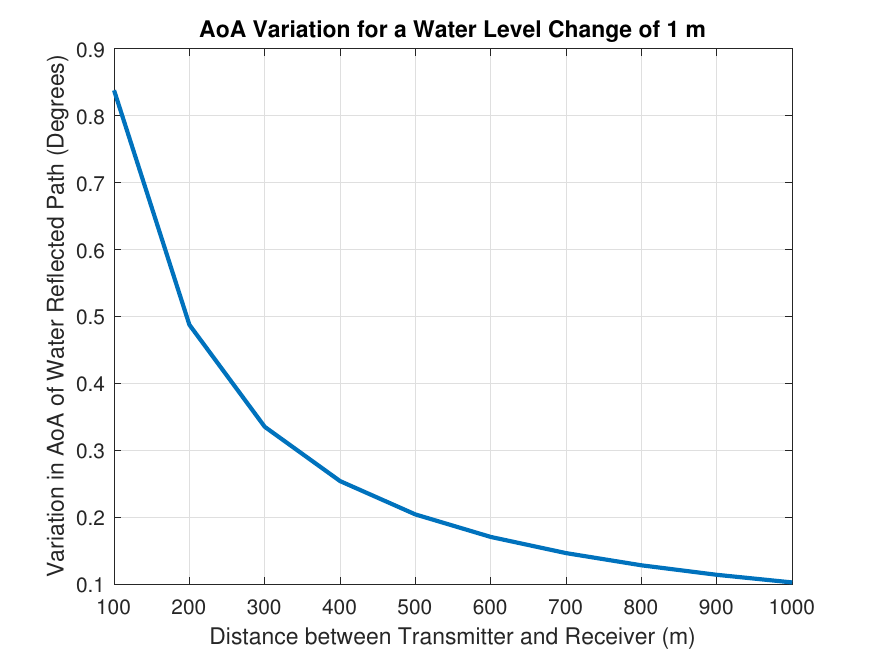} 
  \caption{Water Reflected Path AoA variations vs Distance between Tx and Rx. The total water level change is $1$ $m$.}
\label{fig:AoAvariation}
\end{figure}

As a result of the above analysis, the change in phase is mainly due to the variation in path length. This can be modeled as
\begin{align}
  \psi\left(t\right) = 2\pi d_1\left(t\right)/\lambda,
  \label{phasechange}
\end{align}
where $\psi\left(t\right)$ denotes the phase of the water-reflected path at time $t$. As the water level increases or decreases, $d_1\left(t\right)$ changes and, consequently, $\psi\left(t\right)$ changes. {The beamformed CSI in (\ref{BFCSI}) can be written as
\begin{align}
  \mathbf{h}_{BF}=&\mathbf{w}_1^H \mathbf{H}_l \nonumber \\
    =&\mathbf{w}_1^H \left[\mathbf{h}_l\left(t_0\right),\mathbf{h}_l\left(t_0\right),\dots \mathbf{h}_l\left(t_{N-1}\right)\right], 
    \label{eq:BFCSI}
\end{align}
where $\mathbf{h}_l\left(t\right)=\mathbf{a}\left(\theta_1\left(t\right)\right) \alpha_1\left(t\right)e^{-j2\pi \left(d_1\left(t\right)/\lambda\right)}+\mathbf{n}_l\left(t\right)$, with $\mathbf{n}_l\left(t\right)$ being the combination of the noise and residual from other paths. Based on the AoA change assumption validated in Fig. \ref{fig:AoAvariation}  and ignoring the path gains, variation over time, we can write
\begin{align}
  \mathbf{h}_l\left(t\right)=\mathbf{a}\left(\theta_1\right) \alpha_1e^{-j2\pi \left(d_1\left(t\right)/\lambda\right)}+\mathbf{n}_l\left(t\right),
\end{align}
where $\mathbf{a}\left(\theta_1\right)\approx\mathbf{a}\left(\theta_1\left(t\right)\right)$ and $\alpha_1\left(t\right)\approx\alpha_1$ are applied.
Then the BF CSI can be written as
\begin{align}
      &\mathbf{h}_{BF}=\mathbf{w}_{lcmv}^H \left(\mathbf{a}\left(\theta_1\right)\alpha_1\left[e^{-j2\pi \left(d_1\left(t_0\right)/\lambda\right)},\right. \right. \nonumber \\ & \left.\left. e^{-j2\pi \left(d_1\left(t_1\right)/\lambda\right)}, \dots, e^{-j2\pi \left(d_1\left(t_{N-1}\right)/\lambda\right)}\right] +\mathbf{N}_l\right) \nonumber \\
            &\approx \hat{\alpha}_1\left[e^{-j2\pi \left(d_1\left(t_0\right)/\lambda\right)},  \dots, e^{-j2\pi \left(d_1\left(t_{N-1}\right)/\lambda\right)}\right]. %
\end{align}
Note that $\mathbf{w}_1^H \mathbf{a}\left(\theta_1\right)\approx1$, provided that $\theta_1 $ is correctly estimated. Moreover, $\mathbf{N}_l$ represents the noise and residuals of other paths over time.} Then the phase variations of $\mathbf{h}_{BF}$ are caused by the change in water-reflected path length. Therefore, the phase of $\mathbf{h}_{BF}$ is first calculated using the following,
\begin{align}
  \hat{\Psi}=&\angle\left(\mathbf{h}_{BF}\right) \nonumber \\
  =&\frac{-2\pi}{\lambda}[\hat{d}_1\left(t_0\right), \hat{d}_1\left(t_1\right), \dots, \hat{d}_1\left(t_{N-1}\right)] \nonumber\\
  =&\frac{-2\pi}{\lambda} \Delta\hat{\mathbf{d}},
\end{align}
where $\Delta \hat{\mathbf{d}}=[\hat{d}_1\left(t_0\right), \hat{d}_1\left(t_1\right), \dots, \hat{d}_1\left(t_{N-1}\right)]$ is the estimated path length change between time $t_0$ and $t_{N-1}$. Here, the estimated path length at time $t$, namely $\hat{d}\left(t\right)$, does not represent the absolute path length of the water-reflected signal due to the periodicity of the phase. However, the relative path length with respect to the initial path length at any time $t$, $\Delta d\left(t\right)= d\left(t\right)-d\left(t_{0}\right)$ can be accurately estimated. When phase variations over time $T$ are continuously captured, the phase periodicity does not add any ambiguity to the relative path length. Moreover, the unwrapped phase represents the relative variation in phase. The path length can be calculated as 
\begin{align}
  \Delta \hat{\mathbf{d}}=&\frac{-\hat{\Psi} \lambda}{2\pi},
  \label{pathlenght}
\end{align}
where the negative sign indicating slopes of phase change and path length change have opposite directions. 

\subsection{Water Level Change Estimation}
{In Fig. \ref{fig:FlowDiagram}, the final block is water level change estimation, requiring $\Delta\mathbf{d}$ and water reflection angle as inputs.}
To do this, we need to utilize the geometry of the Tx-Rx setup. Specifically, the AoA of the water-reflected path is used to obtain the incident/reflected AoA from the water surface. We Assume a perfect reflection from the water surface, i.e., $\alpha_i\left(t\right)=\alpha_r\left(t\right)=\alpha\left(t\right)$, as illustrated in Fig. \ref{fig:SysM}, to simplify the modeling. In practice, minor deviations from perfect reflection may exist due to surface roughness or wind. However, our method remains robust, as it leverages CSI samples within a long time window to extract stable AoA features while estimating water level variations. We also adopt a practical assumption that the water level only causes a very small angle change, as explained earlier in Fig. \ref{fig:AoAvariation}. These conditions lead to $\alpha\left(t\right)=\alpha$. Moreover, as the AoA of the water-reflected path $\hat{\theta}_1$ has been estimated using the 2D spectrum estimation, we can calculate the water reflection angle as $\hat{\alpha}=\theta_{inc}-\hat{\theta}_1$, where $\theta_{inc}$ is the inclination angle of the Rx array with respect to the vertical axis; see Fig. \ref{fig:SysM}.
\begin{figure}[!b]
\centering
  \includegraphics[width=\columnwidth]{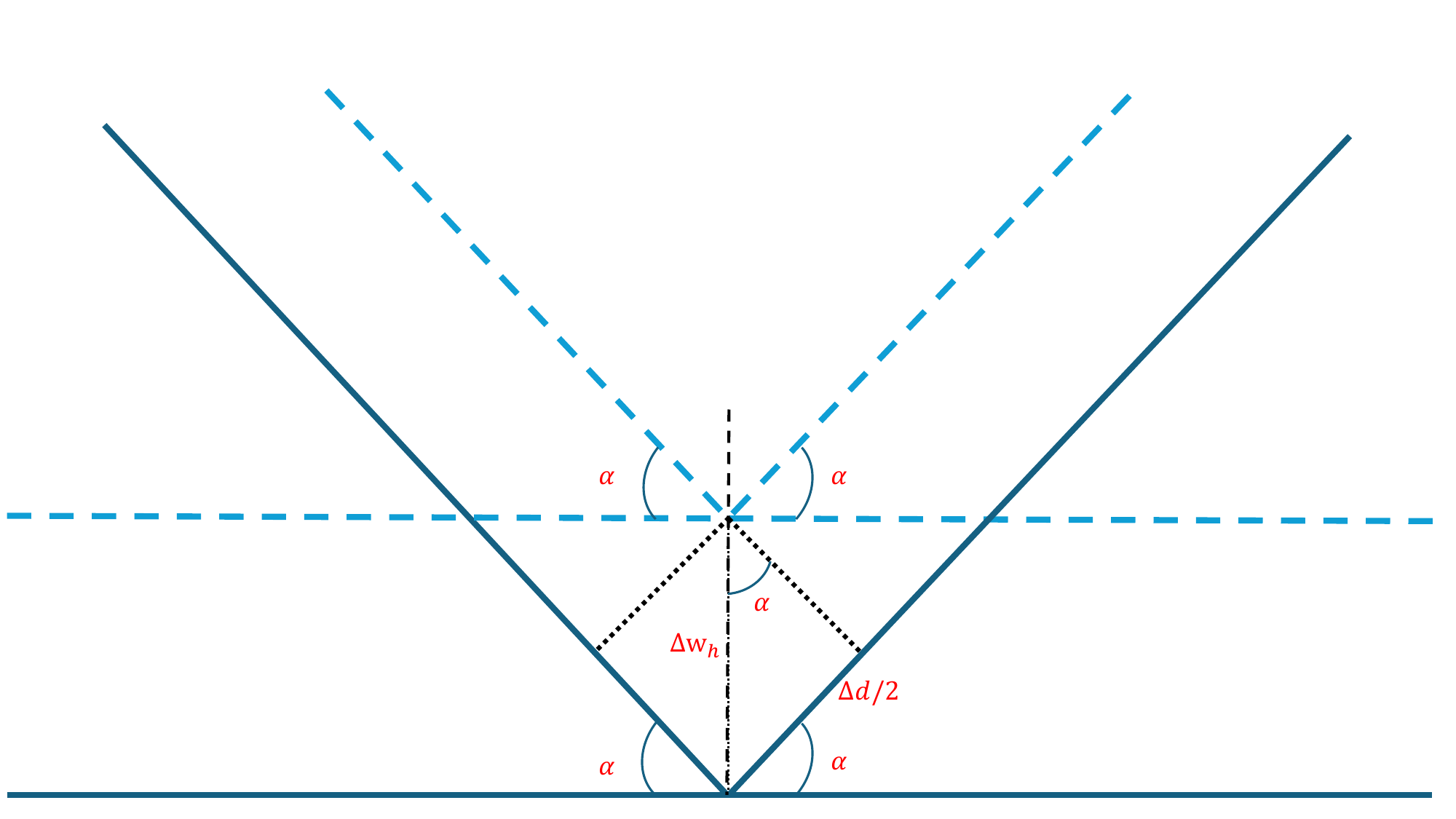} 
  \caption{From path length change to water level change.}
  \label{fig:WaterLevel1}
\end{figure}

Since the reflection angle is assumed to vary negligibly, we can calculate the change in water level from the change in path length based on Fig. \ref{fig:WaterLevel1}. Here, Fig. \ref{fig:WaterLevel1} is the zoomed-in version of the geometry shown in Fig.\ref{fig:SysM}, based on the approximation of $\alpha$. For a total path length change of $\Delta {{d}}=d\left(t_{N-1}\right)-d\left(t_0\right)$, $\Delta {{d}}/2$ is the change in the path length for Tx to water, and the same is the case from water to the Rx. Then using the triangle on the right we can write 
\begin{align}
  \sin \left({\alpha}\right)=\frac{\Delta {{d}}/2}{\Delta {{w}}_h},
\end{align}
where $\Delta w_h= w_h\left(t_{N-1}\right)- w_h\left(t_0\right)$. Similarly, we can write the instantaneous change in water level in terms of instantaneous change in water-reflected path length as
\begin{align}
  \Delta \hat{\mathbf{w}}_h=\frac{-\Delta \hat{\mathbf{d}}/2}{\sin\left(\hat{\alpha}\right)}= \frac{-\Delta \hat{\mathbf{d}}}{2 \sin\left(\hat{\alpha}\right)},
  \label{WaterlevelChange}
\end{align}
where $\Delta \hat{\mathbf{w}}_h=[w_h\left(t_0\right),w_h\left(t_1\right),\dots, w_h\left(t_{N-1}\right)]$. The negative sign indicates that an increase in path length translates into a decrease in water level, and vice versa. Based on the above, if the water reflection angle and the change in path length are correctly estimated, the change in water level can be estimated. 

\section{Experimental Results and Discussion}

In this section, field experimental results are presented to demonstrate the performance of the proposed methods in practical scenarios. There are three different datasets collected infield on different days. The BS transmits LTE signals at the carrier frequency $f_c=2659.8$ MHz and a bandwidth of $20$ MHz. The Rx is based on the Xilinx ZC706 evaluation board \cite{ZC} which captures the downlink LTE signals with a baseband sampling rate of $30.72$ MHz. The AD-FMCOMM55-EZB is the RF board housed by the ZC706 evaluation board. This RF board has two AD9361 transceivers \cite{ad}. Each AD9361 has two fully synchronized transceivers.  

\begin{figure}[!b]
\centering
  \includegraphics[width=0.45\textwidth]{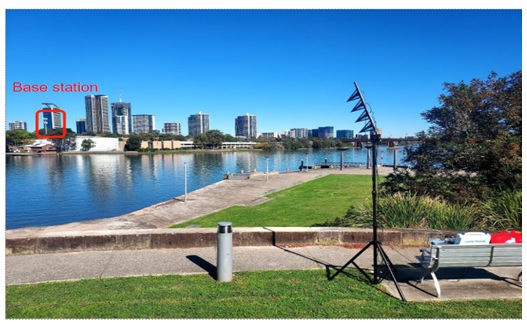} 
  \caption{Setup 3, Capturing LTE Signals From the Base Station. Location: Meadowbank, NSW, Australia, Base Station Coordinates: Latitude $-33.825275^{o}$, Longitude $151.091715^{o}$ (WGS84).}
  \label{fig:DataCollectionSetup}
\end{figure}
\subsection{Receiver Array Calibration}
\begin{figure*}[!t] %
  \centering %
  \begin{subfigure}{0.3\textwidth} %
    \centering
    \includegraphics[width=\linewidth]{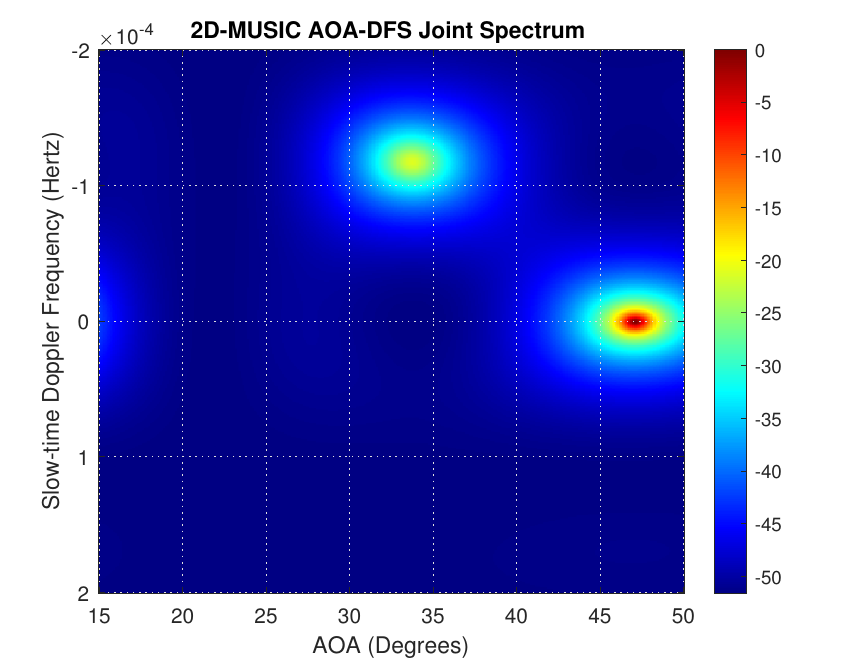} %
    \caption{Setup 1} %
    \label{fig:subfig1} %
  \end{subfigure}
  \begin{subfigure}{0.3\textwidth} %
    \centering
    \includegraphics[width=\linewidth]{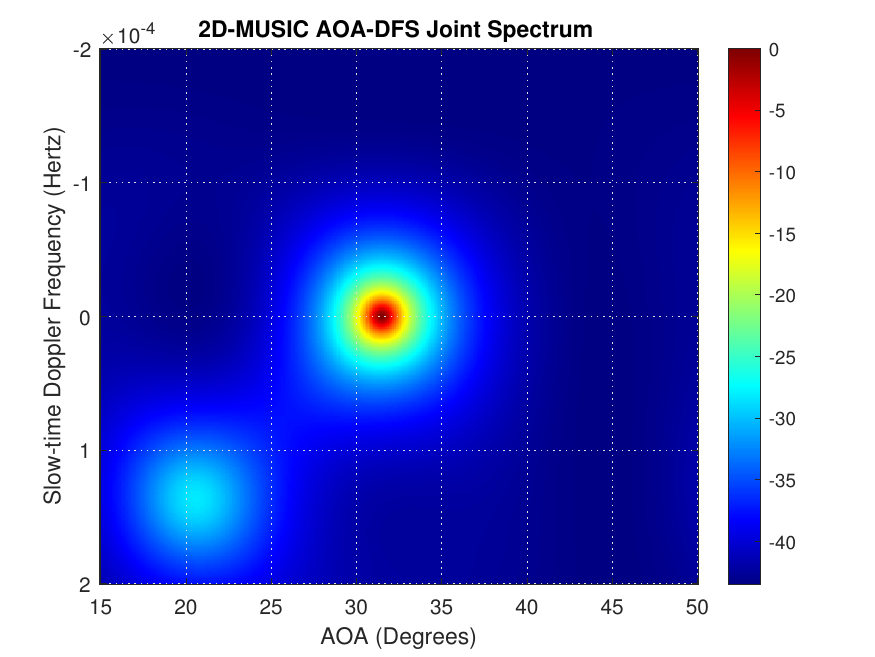} %
    \caption{Setup 2} %
    \label{fig:subfig2} %
  \end{subfigure}
  \begin{subfigure}{0.3\textwidth} %
    \centering
    \includegraphics[width=\linewidth]{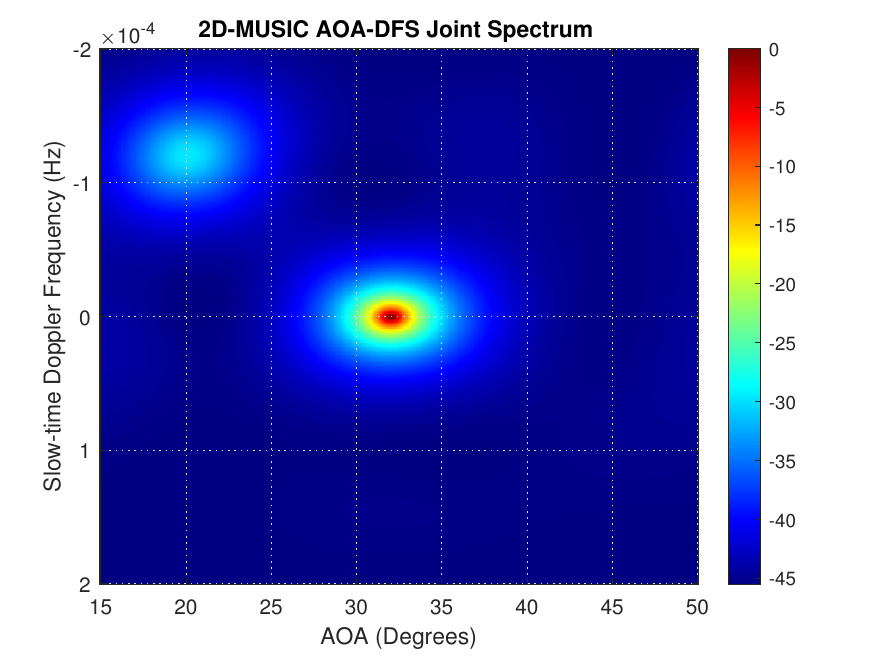}
        \caption{Setup 3} %
    \label{fig:subfig2} %
  \end{subfigure}
  \caption{Joint estimate of AoA-Slow-time DFS, LOS path with zero DFS and water-reflected path with non-zero DFS.} %
  \label{fig:2Dspectrum} %
\end{figure*}

\begin{figure*}[!t] %
  \centering %
  \begin{subfigure}{0.3\textwidth} %
    \centering
    \includegraphics[width=\linewidth]{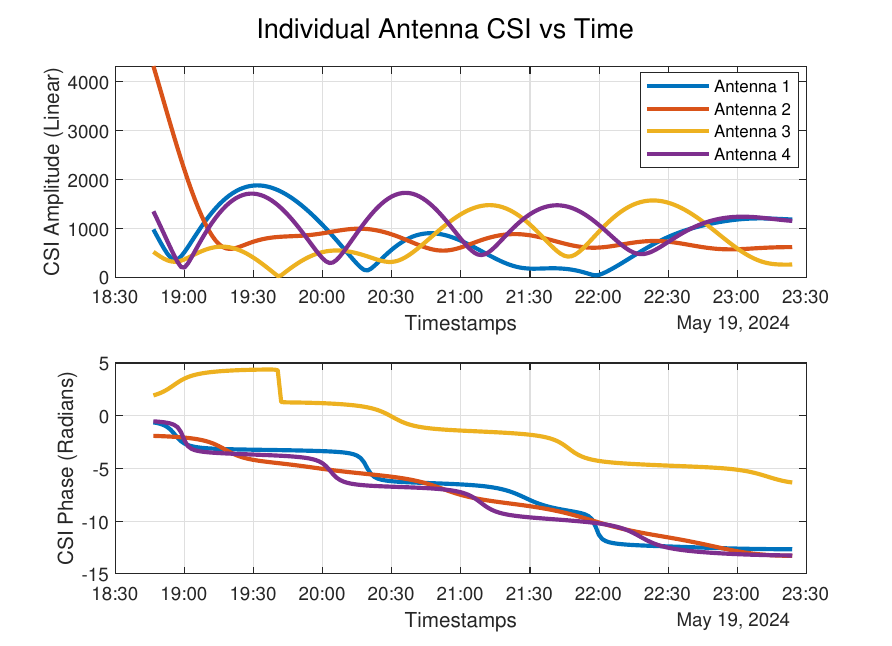} %
    \caption{Setup 1} %
    \label{fig:subfig1} %
  \end{subfigure}
  \begin{subfigure}{0.3\textwidth} %
    \centering
    \includegraphics[width=\linewidth]{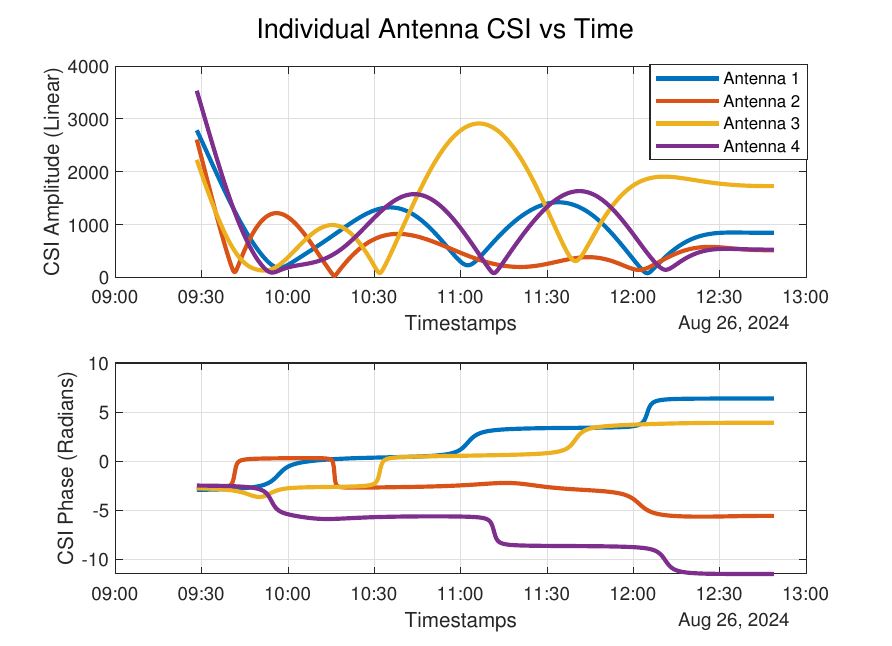} %
    \caption{Setup 2} %
    \label{fig:subfig2} %
  \end{subfigure}
  \begin{subfigure}{0.3\textwidth} %
    \centering
    \includegraphics[width=\linewidth]{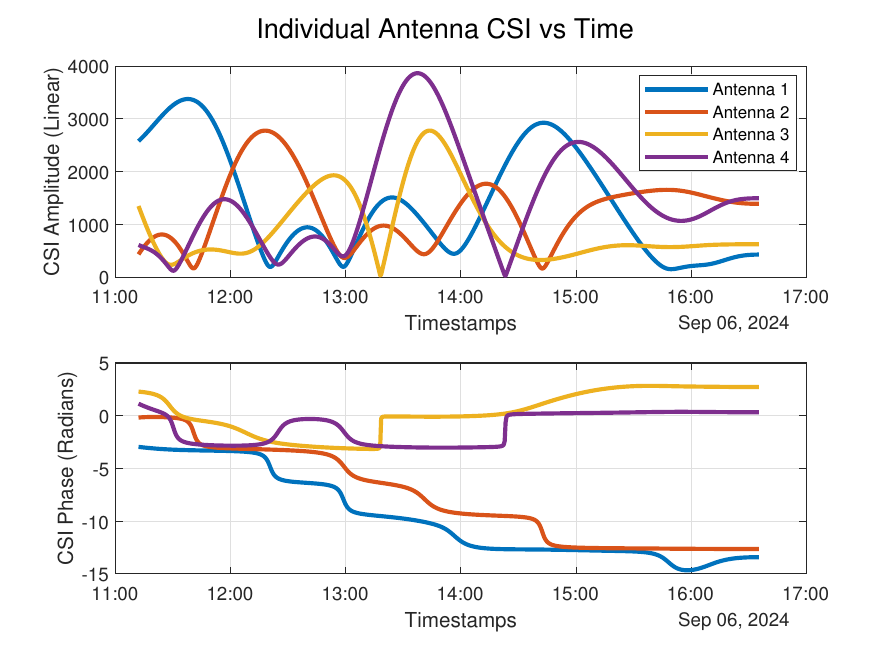}
        \caption{Setup 3} %
    \label{fig:subfig2} %
  \end{subfigure}
  \caption{CSI amplitude and phase over time on individual antennas.} %
  \label{fig:CSIInd} %
\end{figure*}
\begin{figure*}[ht] %
  \centering %
  \begin{subfigure}{0.3\textwidth} %
    \centering
    \includegraphics[width=\linewidth]{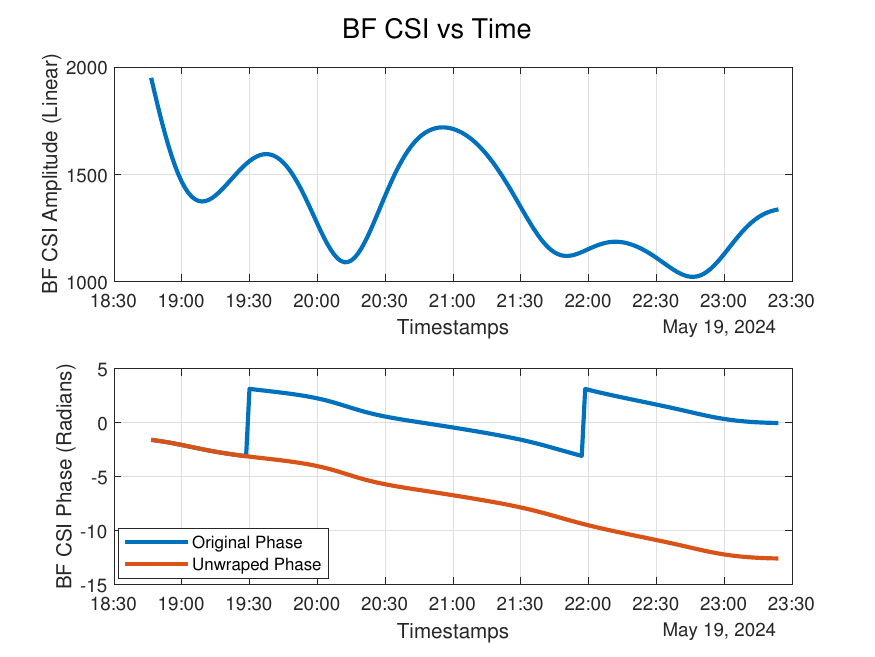} %
    \caption{Setup 1} %
    \label{fig:subfig1} %
  \end{subfigure}
  \begin{subfigure}{0.3\textwidth} %
    \centering
    \includegraphics[width=\linewidth]{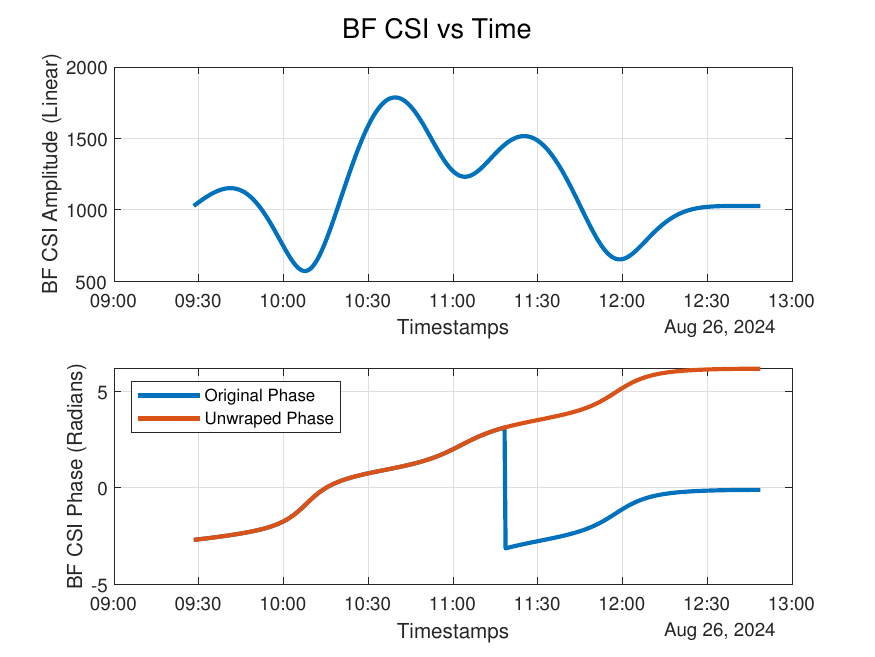} %
    \caption{Setup 2} %
    \label{fig:subfig2} %
  \end{subfigure}
  \begin{subfigure}{0.3\textwidth} %
    \centering
    \includegraphics[width=\linewidth]{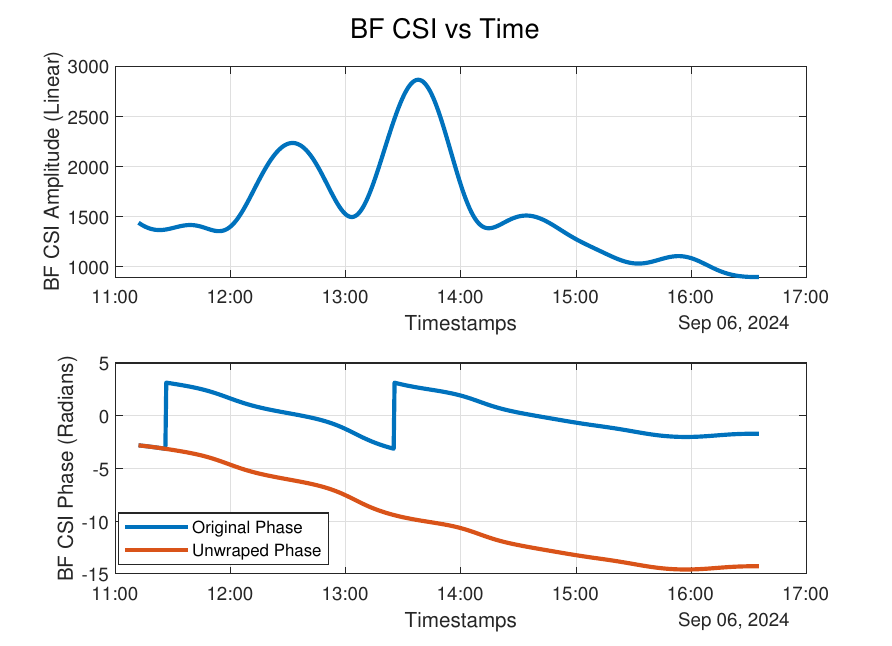}
        \caption{Setup 3} %
    \label{fig:subfig2} %
  \end{subfigure}
  \caption{BF CSI amplitude and phase over time using LCMV beamformer weights.} %
  \label{fig:BFCSI} %
\end{figure*}
The two pairs of AD9361 in AD-FMCOMMS5-EBZ are not fully synchronized due to the use of individual local oscillators in each AD9361. Channels 1,2 and channels 3,4 are fully synchronized, and there is a time-invariant random clock offset (RCO) between the two pairs of channels. This RCO is dependent on the $f_c$ and sampling rate and needs to be calibrated as the $f_c$ or sampling rate is changed. This RCO can cause ambiguity in spatial domain processing resulting in incorrect AoA estimation. In addition to the RCO, antenna gains and phase errors can also cause ambiguity in the AoA estimation. In order to tackle these issues, array calibration based on the pilot target is implemented in this work. Knowing the location of the Tx and Rx, the AoA of the LOS can be calculated based on the geometry. Then using the calculated LOS AoA, namely $\theta_0$, the combined array errors caused by the antenna gain/phase errors and RCO can be collectively compensated. 
 
For array calibration, raw baseband signals are used before performing the downlink channel estimation.  A subspace-based calibration method \cite{soon1994subspace} is used for array calibration. The collective effect of antenna errors and RCO can be modeled as 
 \begin{align}
   &\mathbf{E}=\text{diag}\left(\mathbf{e}\right)\\\nonumber
   &=\text{diag}\left(\left[1, \zeta_1 e^{-j\beta_{1}}, \zeta_2 e^{-j\left(\beta_{2}+\beta_{c}\right)}, \zeta_3 e^{-j\left(\beta_{3}+\beta_{c}\right)}\right]\right),
 \end{align} 
where $\zeta_i$ and $\beta_{i}$ denote the gain error and phase error, respectively. They are caused by antenna impairments for the $i$-th antenna. In addition, $\beta_{c}$ denotes the RCO between the two AD9361 transceivers with the first transceiver used as a reference with $\beta_c=0$. Thus, we can express the baseband received signal as 
 \begin{align}
 \mathbf{B}=\mathbf{E}\mathbf{A} \mathbf{S}+ \mathbf{N}
 \in \mathbb{C}^{M \times G},   
 \end{align}
 where $G$ denotes the number of time domain samples, $\mathbf{S}$ is the source matrix signal and $\mathbf{N}$ is the noise matrix. Furthermore, $\mathbf{A}=[\mathbf{a}\left(\theta_0\right),\mathbf{a}\left(\theta_1\right),\dots, \mathbf{a}\left(\theta_{S-1}\right) ]$ denotes the array matrix with $S$ sources. The covariance matrix can be written as
 \begin{align}
   \mathcal{R}=\mathbf{B}\mathbf{B}^H=\mathbf{E} \mathbf{A} \mathbf{R}_s \mathbf{A}^H \mathbf{E}^H+ \sigma^2 \mathbf{I},
 \end{align}
where $\mathbf{R}_s$ is the source covariance matrix and $\sigma^2$ is the noise variance. After performing the EVD of $\mathcal{R}$ we have,
\begin{align}
  \tilde{\mathcal{R}}= \mathbf{E_s} \mathbf{\Omega} \mathbf{E_s}^H + \sigma^2 \mathbf{E}_N \mathbf{E_N},
\end{align}
with $\mathbf{E}_s$ as the eigen vector spanned by $\mathbf{E}\mathbf{A}$. If $\mathbf{A}$ is known, that is, the array response of the pilot target is known, the array errors can be estimated using the method in \cite{soon1994subspace}. 
In our case, the LOS path can serve as a pilot target with the known AoA given by $\theta_0$. As such, the EVD of the matrix, $\mathbf{A}_0 ^H \mathbf{E_s}\mathbf{E_s}^H \mathbf{A}_0$ is calculated, where $\mathbf{A}_0=\text{diag}\left(\mathbf{a}\left(\theta_0\right)\right)$. Then the eigen vector that has a unity eigen value is considered as the estimated array impairments $\hat{\mathbf{e}}$ \cite{soon1994subspace}.

\subsection{Field Experiment Results}
\label{sec:FieldExp}
To demonstrate the performance of the proposed scheme, three different setups were used to capture the LTE signals from the BS. The detailed list of parameters for three setups corresponding to the parameters defined in Fig. \ref{fig:SysM} is given in Table \ref{tab:Setups}. Setup 1 and Setup 3 are the same except for the inclination angle of the Rx array. For Setup 2, the Rx array is moved farther from the river's bank and the total distance between Tx and Rx is increased. The sampling rate $f_s=30.72$ MHz, carrier frequency $f_c=2659.8$ MHz, bandwidth $B=20$ MHz. Each CSI capture consists of $K=200$ sub-carriers with sub-carrier spacing of $90$ kHz, $N=200$ symbols with symbol rate of 2000 symbols/sec. As an example, Setup 3 is shown in Fig. \ref{fig:DataCollectionSetup}.

\begin{figure*}[ht] %
  \centering %
  \begin{subfigure}{0.3\textwidth} %
    \centering
    \includegraphics[width=\linewidth]{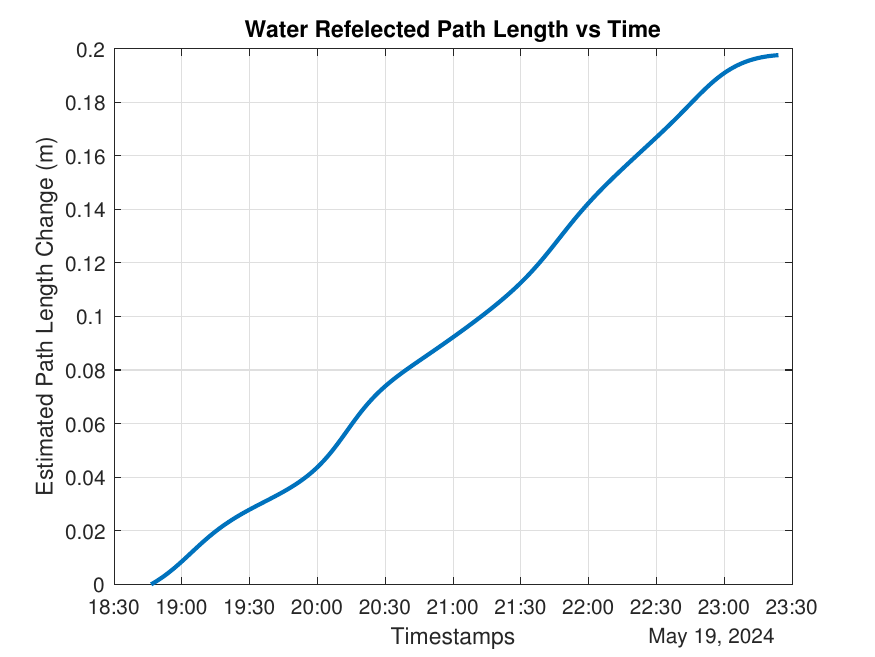} %
    \caption{Setup 1} %
    \label{fig:subfig1} %
  \end{subfigure}
  \begin{subfigure}{0.3\textwidth} %
    \centering
    \includegraphics[width=\linewidth]{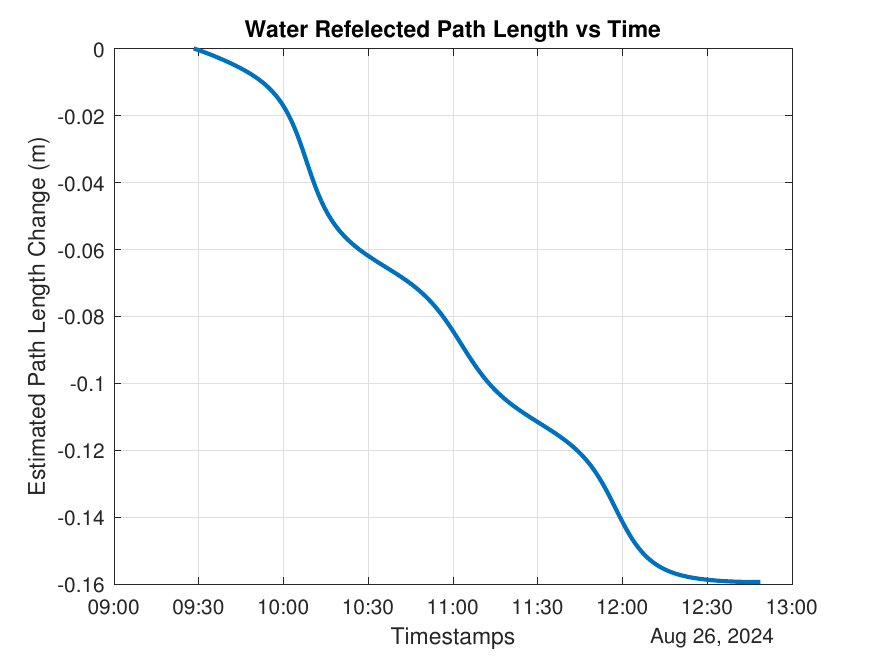} %
    \caption{Setup 2} %
    \label{fig:subfig2} %
  \end{subfigure}
  \begin{subfigure}{0.3\textwidth} %
    \centering
    \includegraphics[width=\linewidth]{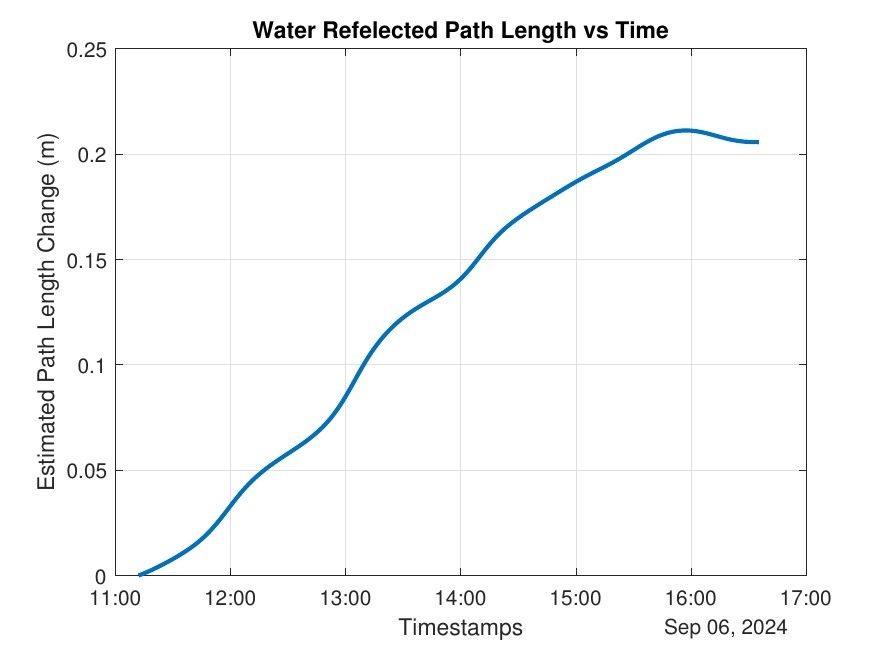}
        \caption{Setup 3} %
    \label{fig:subfig2} %
  \end{subfigure}
  \caption{Path length Change over time using the extracted phase variations.} %
  \label{fig:Pathlength} %
\end{figure*}

\begin{figure*}[!t] %
  \centering %
  \begin{subfigure}{0.3\textwidth} %
    \centering
    \includegraphics[width=\linewidth]{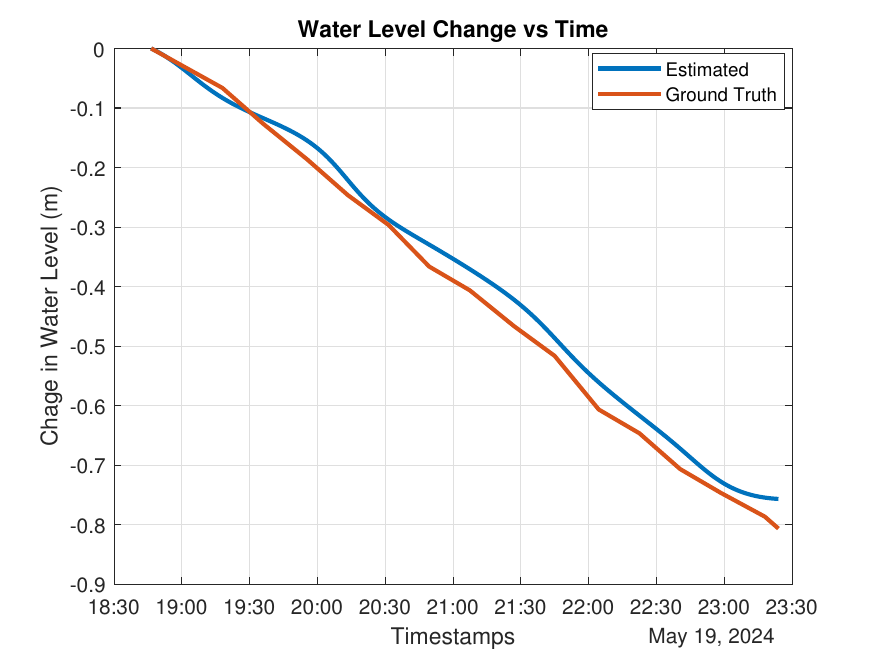} %
    \caption{Setup 1} %
    \label{fig:subfig1} %
  \end{subfigure}
  \begin{subfigure}{0.3\textwidth} %
    \centering
    \includegraphics[width=\linewidth]{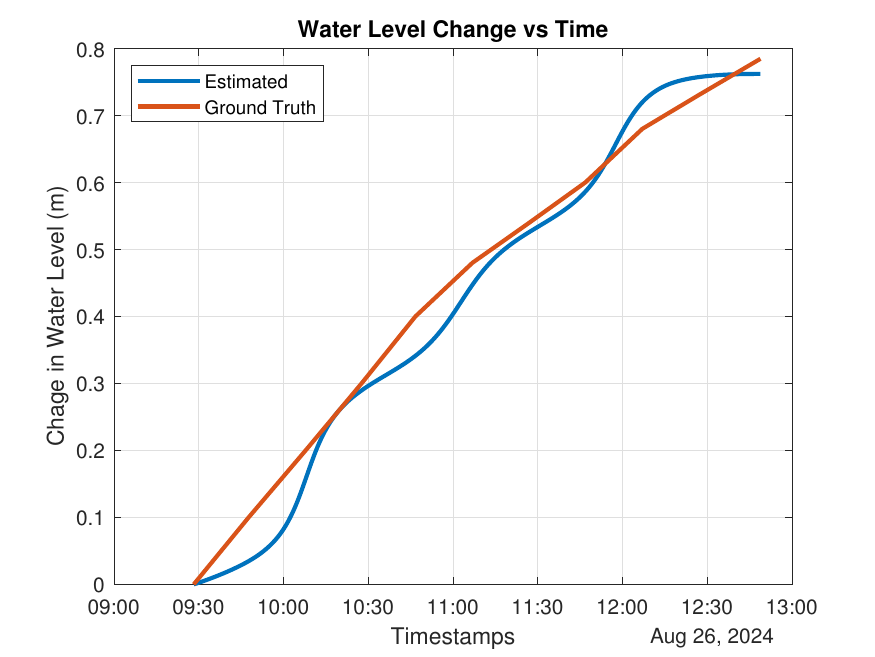} %
    \caption{Setup 2} %
    \label{fig:subfig2} %
  \end{subfigure}
  \begin{subfigure}{0.3\textwidth} %
    \centering
    \includegraphics[width=\linewidth]{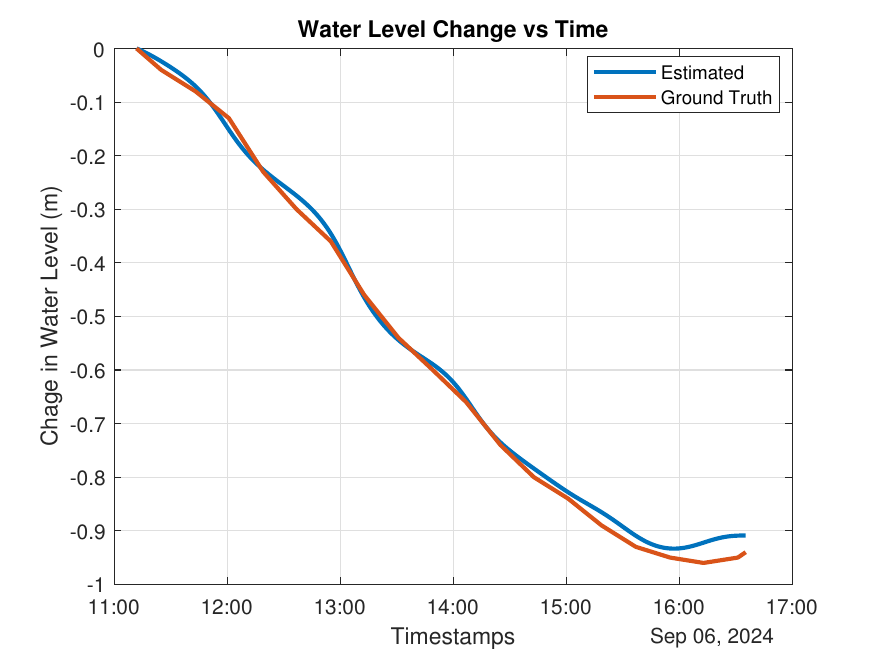}
        \caption{Setup 3} %
    \label{fig:subfig2} %
  \end{subfigure}
  \caption{Water level change over time using the path length change estimate.} %
  \label{fig:WaterLevel} %
\end{figure*}

\begin{table}
  \centering
  \caption{Data Collection Setups}
\begin{tabular}{|c|c|c|c|} \hline 
     Parameters & Setup 1 & Setup 2& Setup 3 \\ \hline 
 Date & 19- May-24& 26-Aug-24& 6-Sep-24\\\hline \hline 
     $D_{tr}$& $423$ m& $465$ m& $423$ m\\ \hline 
     $D_{rw}$& $0.75$ m& $34$ m& $0.75$ m\\ \hline 
     $D_{tw}$& $160$ m& $151$ m& $160$ m\\ \hline 
     $H_t$& $45$ m& $45$ m& $45$ m\\ \hline 
     $H_r$& $4$ m& $10$ m& $4$ m\\ \hline 
     $\theta_{inc}$& $42^\circ $& $27 ^\circ $& $27 ^\circ $\\\hline
 $\Delta t$& 90 sec& 28 sec&22 sec\\\hline
 $T$& $4.5$ hours & $3.25$ hours &$5.25$ hours \\\hline
  \end{tabular}
\label{tab:Setups} 
\end{table}
After the RPO compensation in Section \ref{sec:RPO} and CSI dimension reduction in Section \ref{sec:DimRed}, the next step is to estimate the AoA of the water-reflected path using joint space-time processing proposed in Section \ref{sec:AoAEstimate}. 
In the joint space-time heatmap in Fig. \ref{fig:2Dspectrum}, two paths can be seen for each setup; the path with zero Doppler is the LOS path and that with non-zero Doppler is the water-reflected path. The AoAs of the  water-reflected paths,  $\hat{\theta}_1$ are estimated as  $34.5 ^\circ$, $21 ^\circ$, $20 ^\circ$ for Setups 1, 2, and 3, respectively. The sign of the slow-time DFS is related to the direction of the water level change. For Setup 1 and Setup 3 the slow-time DFS is negative, indicating a decrease in  the water level. For Setup 2, the slow-time Doppler is positive, indicating an increase in the water level.

Once the water-reflected path's AoA, $\theta_1$, has been successfully estimated, the next step is to remove the static clutter, including the LOS path, by subtracting the long-term temporal mean using (\ref{eq:clutter}). In Fig. \ref{fig:CSIInd}, the CSI amplitude and phase over time are shown. The phase variation trend can be seen for all setups, but still, it is hard to get a clear phase variation trend from individual antenna's CSI due to the residue from interfering paths. For Setups 1 and 3, the phase has a decreasing trend, and for Setup 2, the phase has an increasing trend on most antennas. %

To obtain a clear phase variation, the LCMV beamforming weights are calculated based on (\ref{eq:BFWeights}). The desired direction is the water reflected path's AoA $\theta_1$ and the null direction is the LOS path's  $\theta_0$. Using the LCMV weights, the BF CSI is extracted using (\ref{eq:BFCSI}). Although the LOS path is greatly suppressed by subtracting the long-term temporal mean, this nulling here further suppresses any residue from the LOS path. In Fig. \ref{fig:BFCSI}, the phase variation over time can be clearly seen. For Setups 1 and 3, the phase is decreasing over time, indicating an increase in path length of the water-reflected path. Whereas, for Setup 2, the phase is increasing, indicating a decrease in path length of the water-reflected path.

The estimated phase variations over time are then converted to water-reflected path length change using (\ref{pathlenght}). The initial path length is assumed to be zero. In Fig. \ref{fig:Pathlength} it can be seen that the path length is increasing for Setups 1 and 3 and decreasing for Setup 2, where the decrease in path length indicates an increase in water level and an increase in path length indicates a decrease in water level. 

Finally, the change in the path length of the water-reflected path is then translated to change in water level using (\ref{WaterlevelChange}). In Fig. \ref{fig:WaterLevel}, for Setups 1 and 3, the water level is decreasing and can be seen matching the ground truth accurately. For Setup 2, the water is increasing and the estimated and ground truth are matching well. For Setup 2, the Rx is moved $34$ m farther from the river bank; still, the sensing performance shows minimal degradation. The mean water level estimation error for Setup 1 is $2.62$ cm with a standard deviation of $1.27$ cm. Similarly, for Setup 2, the mean estimation error and standard deviation are $3.05$ cm and $2.16$ cm, respectively. Finally, the mean estimation error and standard deviation for Setup 3 are  $1.5$ cm and $1.01$ cm, respectively. The ground truth is obtained using the data at \cite{WaterLevelGT}.

\section{Conclusions}
 This paper presents a novel passive water level sensing scheme using downlink cellular signals, offering a cost-effective and infrastructure-free alternative to traditional sensor-based monitoring systems. By leveraging existing cellular infrastructure and exploiting variations in channel state information, the proposed method accurately estimates water level changes through joint space-time processing. A beamforming-based technique is introduced to enhance the water-reflected path, while a compensation scheme addresses clock asynchronism between the transmitter and receiver. Experimental results from a practical river scenario validate the effectiveness of the proposed approach, showing high agreement with ground truth measurements.
The proposed sensing scheme also shows strong potential for integration into mobile infrastructure, enabling uplink CSI to be utilized for high-accuracy water sensing. This paves the way for scalable and resilient environmental monitoring, particularly in flood-prone or infrastructure-limited regions.

\section{Acknowledgment}
We thank Dr Ayoob Salari for assistance with experimental data collection. We also gratefully acknowledge the NSW State Emergency Service for their valuable feedback as part of a 5G flood sensing project related to this research.

\bibliographystyle{IEEEtran}
\bibliography{Reference}
\end{document}